\documentclass{aa}  
\usepackage{tikz}
\usepackage{graphicx}
\usepackage{txfonts}
\usepackage[colorlinks=true,linkcolor=red, citecolor=blue]{hyperref}
\usepackage{fontawesome}
\usepackage{amssymb}
\usepackage{mathtools}
\usepackage{lipsum}
\usepackage{comment}
\usepackage{subcaption}         % necessary for continued figures, example in section 3
\usepackage{lscape}             % to rotate a single page table, example in appendix.
\usepackage{placeins}           % useful with \FloatBarrier, to keep 
\newcommand{\td}{\textrm{d}}
\newcommand{\mh}{M_{\rm h}}
\makeatletter\let\linenumbers\relax\makeatother
\begin{document}

%%%%%%%%%%%%%%%%%%%%%%%%%%%%%%%%%%%%%%%%
% if you use custom commands in your title,
% ensure to check your title when submitting!
%%%%%%%%%%%%%%%%%%%%%%%%%%%%%%%%%%%%%%%%
    \titlerunning{A universal language for JWST-era EoR galaxies}
  \title{Interpreting JWST UV luminosity functions and angular correlation functions via a ``universal language'' for high-redshift galaxies}
% or \title{"Model-independent" constraints on high-redshift galaxies from JWST}

%%%%%%%%%%%%%%%%%%%%%%%%%%%%%%%%%%%%%%%%
% Please separate each author with the \and command
%
% Please do not include ORCIDs next to author names.
% Only ORCIDs authenticated by individual authors in EDPS
% editorial system will be taken into account.
% ORCIDs included here will be removed.
%%%%%%%%%%%%%%%%%%%%%%%%%%%%%%%%%%%%%%%%

   \author{Ivan Nikolić
          \inst{1,2,3}
          \and
          Andrei Mesinger
          \inst{4, 3}
          \and
          Steven G. Murray
          \inst{5, 3}
}

   \institute{
              Cosmic Dawn Center (DAWN), Denmark
             \and
             Niels Bohr Institute, University of Copenhagen, Jagtvej 128, 2200 Copenhagen N, Denmark
            \and
            Scuola Normale Superiore, Piazza dei Cavalieri 7, 56125 Pisa, PI, Italy
           \and
            Department of Physics and Astronomy {\it ``Ettore Majorana''}, University of Catania, Via Santa Sofia 64, 95123  Catania, Italy
             \and
             Department of Physics, Stellenbosch University, Matieland 7602, South Africa
        }

   \date{Received; accepted}

% \abstract{}{}{}{}{}
% 5 {} token are mandatory
 
  \abstract
   {The James Webb Space Telescope (JWST) found a higher abundance of bright $z\gtrsim10$ galaxies than expected before its launch. No consensus has emerged on whether the difficulties in reproducing these abundances were due to missing physics, simplifying assumptions that break down at the highest redshifts, or model mis-calibration.
    This is complicated because most galaxy models cannot be compared like-to-like, and models invoking very different physical processes can make similar predictions for compressed summaries such as the ultraviolet luminosity functions (UV LFs).
    %
    % [draft placeholder, kept as a comment] Here we take a data-driven approach... simulation-based priors
    Here we describe galaxy populations in a ``universal language'': a chain of physically and empirically motivated conditional distributions linking halo mass, stellar mass, star formation rate and UV magnitude, defined consistently across models and data. We build a broad prior from fits to five hydrodynamic simulations, which predict very different high-redshift galaxy populations.  %, despite the fact that several result in similar observable UV LFs.  
    We infer the parameters of this flexible galaxy model from $z>6$ JWST observations of UV LFs and angular correlation functions (ACFs). 
    %Expressed in this language, the five simulations predict very different galaxy populations, and the inferred population lies within their range.
   % We quantify the relative roles of the different sources of scatter in reproducing the data, rather than describing them with a single scatter in UV magnitude at fixed halo mass.    
Including the mass-complete ACFs disfavors a large scatter around the stellar-to-halo mass relation (SHMR).  
    %, $\sigma_{\rm SHMR}=0.35^{+0.25}_{-0.19}$\,dex, compared with $0.45^{+0.41}_{-0.26}$\,dex from the UV LFs alone. 
    %
    With simulation-based priors, the star-forming main sequence (SFMS) is described by a star-formation timescale of about one halo dynamical time, and the scatter around the SFMS increases towards lower stellar masses.   
    The SFMS scatter is also the main source of stochasticity in UV magnitude at fixed halo mass, accounting for a median $\sim61\%$ of the total variance at $M_{\rm UV}=-20$ (against $\sim$36\% for the SHMR scatter).
    %SFMS scatter is the main source of stochasticity in UV magnitudes at fixed halo mass, with a contribution to the total variance that is on average double than that from SHMR scatter. 
   %Finally, we perform Bayesian model selection with several extensions to our fiducial model, including: (i) adding a redshift evolution of the normalization and slope of the SHMR; (ii) assuming the scatter around the SHMR depends on mass; and (iii) allowing the slope of the SFMS to be a free parameter (instead of being proportional to the halo dynamical time).  In all cases, the resulting Bayes factors are small, meaning that current data does not favor a more complex galaxy model.
Using Bayesian model selection, we find no evidence for redshift evolution of the SHMR or for an SFMS slope different from unity.   
   At $z\sim20$, our fiducial model predicts UV LFs orders of magnitude below estimates from photometric candidates.  
A consensus on UV LF estimates at $z\geq14$, which currently have large uncertainties and disagree, will be critical to determine whether model extensions are required.
   }

   \keywords{ Galaxies: high-redshift -- intergalactic medium -- Cosmology: dark ages, reionization, first stars
               }

   \maketitle

%%%%%%%%%%%%%%%%%%%%%%%%%%%%%%%%%%%%%%%%%%%%%%%%%%%%%%%%%%%%%%

\defcitealias{Nikolic24}{N24}

\section{Introduction}

One of the most surprising early results from the James Webb Space Telescope (JWST) has been an overabundance of bright galaxies at $z\gtrsim10$. While many of the highest redshift candidates are still awaiting spectroscopic confirmation, UV luminosity functions (UV LFs) from Lyman break candidates exceed pre-JWST estimates at the bright end by a wide margin \citep[e.g.,][]{Donnan2023, Castellano2023, Austin2023, Bouwens2023, McLeod2024, Chemerynska2026}.  Proposed explanations include an increased efficiency and/or burstiness of star formation, associated dust-removing outflows, an evolving stellar initial mass function (IMF), or even changes to cosmology \citep[e.g.,][]{Dekel2023, Ferrara2023, Mason2023, Mirocha2023, Munoz2023, Shen2023, Gelli2024, Hutter2025}. While all of these physical scenarios directly impact the galaxy--halo connection, they cannot be distinguished by the UV LF alone.

Recent analyses have thus extended their interpretation beyond the UV LFs. JWST/NIRSpec observations provide high-resolution spectra that are sensitive to the star-formation histories, chemical abundances and baryon distributions of high-redshift galaxies \citep[e.g.,][]{Curti2023, Roberts-Borsani2024, Tang2026}. For example, UV bands around rest-frame $1500$\,\AA\ (used to construct UV LFs), are typically sensitive to star-formation averaged over the past $100$\,Myr (e.g., \citealt{Kennicutt1998}). On the other hand, low-opacity nebular emission lines (obtainable through high-resolution spectra) are sensitive to star formation on shorter timescales, allowing us to measure star-formation ``burstiness'' \citep{Shapley2023, Cole2025, Munoz2026}. Likewise, combining UV LFs with clustering measurements can provide a better handle on which halos host the observed galaxies: a higher star-formation efficiency shifts galaxies of a given brightness into less massive halos, whereas increased stochasticity mixes in galaxies up-scattered from the far more abundant low-mass halos, an effect that is much stronger for bright galaxies hosted by the exponential, high-mass tail of the halo mass function \citep[e.g.,][]{Giavalisco2001, Moster2010, Mirocha2020, Munoz2023, Gelli2024, Shuntov2025, Nikolic2026}.

Despite a rapidly increasing observational sample, the community has yet to agree on a ``concordance'' model for high-redshift galaxy formation.  This is perhaps not surprising, given that the physics of star formation is among the most complex astrophysical problems in cosmology, spanning orders of magnitude in spatial and temporal scales (e.g., \citealt{Krumholz2014}).  We have access to an unprecedented toolkit of theoretical models describing galaxy formation and evolution across redshifts, including hydrodynamic/radiative transfer simulations \citep[e.g.,][]{Xu2016, Dave2019, Lovell2021, Hopkins2023, Garaldi2024, Shaye2026}, analytic and semi-analytic models (SAMs) \citep[e.g.,][]{Mutch2013, Dayal2014, Sommerville2015, Mirocha2017, Henriques2020, Schneider2021}, as well as empirical models \citep[e.g.,][]{Moster2018, Park2019, Behroozi2020, Nikolic24, Yung2025}. These approaches differ in their complexity, numerical implementation, built-in assumptions, as well as in the quantities they predict.  More fundamentally however, these models \textit{speak different languages}. 
For example, the free parameters which govern a particular implementation of resolution-dependent, sub-grid star formation in a hydrodynamic simulation are difficult to translate to the parameters of a given SAM \citep{Guo2016, Seeyave2025, Mauerhofer2025}. This makes it nearly impossible to compare different models in a like-to-like fashion, even when obtaining identical observables.  As a result, the ``lessons learned'' from observations often only have context in a given model and its associated free parameters.

Alternatively, one could agree on a set of quantities that are defined consistently across all models, such as the star formation rate (SFR), stellar mass ($M_\ast$), halo mass ($M_{\rm h}$), metallicity ($Z_\ast$ or $Z_{\rm gas}$), redshift ($z$), etc. These quantities may be inputs to some models or outputs of others, but either way they are relevant, well-defined, and can act as the interface between models, as well as between models and data. To serve this role, the   relations connecting them need functional forms and priors that capture the range of predictions from existing galaxy formation models, rather than the ad hoc choices common in  empirical approaches. 

Building on previous work \citep[][hereafter \citetalias{Nikolic24}]{Nikolic24}, here we propose just such a universal model, one that is minimalist and flexible, yet physically motivated. Specifically, we model the connection between galaxies and halos through a series of conditional distribution functions motivated by well-established relations, such as the stellar-to-halo mass relation (SHMR) or star-forming main sequence (SFMS).  The parameters characterizing the mean, scatter, redshift evolution of these relations form a sort of ``universal language'', allowing us to use a wide variety of galaxy-formation simulations to motivate functional forms and associated (conservatively-broad) priors.  We find that all five of the simulations we use for this purpose (FLARES, \citealt{Lovell2021}; FIRE, \citealt{Ma2017}; ASTRID, \citealt{Bird2022}; SERRA, \citealt{Pallottini2022}; First Light, \citealt{Ceverino2018}) predict very different galaxy populations, despite several of them having similar predictions for compressed/lower-dimensional quantities such as the UV LF \citep[e.g.,][]{Ceverino2017, Ma2017, Vijayan2021, Bird2022}.

We jointly infer the parameters of our model from the UV LFs of \citet{Willott2024} and the (mass-complete) angular correlation functions (ACFs) of \citet{Paquereau2025}, confirming that our fiducial parametrization is sufficient to recover the two datasets and that they provide complementary information; Appendix~\ref{sec:app_inf_uvlf_others} shows the results for other UV LF estimates \citep{Donnan2024, Harikane2025, Whitler2025}.  We quantify the relative roles of the different sources of scatter in reproducing the data, rather than describing them with a single scatter in UV magnitude at fixed halo mass \citep[e.g.,][]{Mason2023, Shen2023}.
Using Bayesian model selection, we find no evidence for redshift evolution of the SHMR or for an SFMS slope different from unity, and the mean SFMS is well described by a star-formation timescale of about one halo dynamical time.  
We also find that the scatter around the mean SFMS increases towards smaller masses ($89\%$ posterior probability, compared to $83\%$ under the prior), in line with the enhanced low-mass stochasticity predicted by simulations.

This work is structured as follows. In Sect.~\ref{sec:galfor} we describe our galaxy model based on conditional distribution functions. We set up the priors for inference in Sect.~\ref{sec:priors}. We connect the galaxy model to the key observables in Sect.~\ref{sec:observables}. Section~\ref{sec:results} contains the results of the inference, impact on scaling relations and probability distribution within the context of other observations, as well as possible extensions to the model and predictions at $z>14$. We conclude in Sect.~\ref{sec:conclusion}. Throughout, we adopt the Planck 2018 reference cosmology \citep{Planck2020} and the AB magnitude system \citep{Oke1983}.
%%%%%%%%%%%%%%%%%%%%%%%%%%%%%%%%%%%%%%%%%%%%%%%%%%%%%%%%%%%%%%

\section{A "universal language": galaxies as a chain of conditional distributions}
\label{sec:galfor}

In the most general case, {\it any} galaxy population is fully described by a high-dimensional, {\it joint} probability distribution over a given set of properties (e.g., redshift, host halo mass, SFR, stellar mass, metallicity ($Z$), ionizing escape fraction ($f_{\rm ion}$), halo merger history, spin, etc.). Fully characterizing higher-order correlations in such a high-dimensional space is, however, infeasible. Luckily, there are several empirical and physically-motivated correlations that allow us to marginalize over weakly-correlated properties, and describe galaxies with lower-dimensional conditional probabilities whose functional forms are reasonably well established. These conditional PDFs can be characterized with, for example, log-normal distributions whose means and variances depend on $M_\ast$, $z$, SFR, etc.  Both data and models can be compared in the space of these distributions, making them a flexible and interpretable `universal language'.

We illustrate such an approach using the UV luminosity function.  Assuming that all galaxies reside in dark matter (DM) halos\footnote{The halo mass function provides a physical underpinning for all galaxy models as it is well understood (relative to the astrophysics of galaxy formation). Assuming that galaxies are hosted by DM halos allows us to compute their number densities and clustering almost from first principles \citep{Press1974, Sheth1999}.  The challenge then becomes how to assign  galaxies to DM halos.}, we could write the number density of galaxies of a given UV magnitude, $M_{\rm UV}$, at a given redshift, $z$, as
\begin{equation}
\label{eq:UVLFsimple}
\frac{\td n(M_{\rm UV}, z)}{\td M_{\rm UV}} = \int \td\mh \frac{\td n(M_{\rm h}, z)}{\td\mh} p(M_{\rm UV} \mid \mh, z) ~ ,
\end{equation}
where $\td n(M_{\rm h}, z)/\td M_{\rm h}$ is the halo mass function (HMF), and $p(M_{\rm UV} \mid \mh , z)$ is the  conditional probability distribution of a galaxy having UV magnitude $M_{\rm UV}$ if it resides in a halo with mass $M_{\rm h}$ at redshift $z$. A common approximation is to model this distribution as log-normal (i.e., Gaussian in $M_{\rm UV}$), whose mean $\mu_{M_{\rm UV}}(M_{\rm h})$ and standard deviation $\sigma_{\rm UV}$ are then left free to vary. Indeed, interpretations of the bright-end excess in JWST UV LFs often rely on modifying either the mean or the scatter of this distribution \citep[e.g.,][]{Mason2023, Shen2023, Mirocha2023, Munoz2023, Ferrara2023, Sun2023, Hutter2025, Sun2025, Nikolic2026}. However, it is not obvious that $p(M_{\rm UV} \mid M_{\rm h}, z)$ should follow such a simple functional form, as going from $M_{\rm h}$ to $M_{\rm UV}$ is highly complicated, and difficult to express as a product of random variables that would result in a log-normal distribution \citep[see, e.g.,][]{Sun2025, Gelli2026}.  \citealt{Romeo2003}).  
Even if $p(M_{\rm UV} \mid M_{\rm h}, z)$ did follow a log-normal, what priors should be adopted for its means, standard deviations, and how should these evolve with redshift?

Instead, we decompose Eq.~\eqref{eq:UVLFsimple} into a set of conditional relations with well-motivated functional forms and priors.  Specifically, we write:

\begin{equation}
\begin{alignedat}{2}
\frac{\td n(M_{\rm UV}, z)}{\td M_{\rm UV}} = {}& \int_{0}^{\infty} \td\mh\, \frac{\td n(\mh, z)}{\td\mh} &\;& \text{[HMF]}\\
\times {}& \int_0^{f_{\rm b}\mh} \td M_{\ast}\, p(M_{\ast} \mid \mh) && \text{[SHMR]}\\
\times {}& \int_0^{\infty} \td{\rm SFR}_{100}\, p_z({\rm SFR}_{100} \mid M_\ast) && \text{[SFMS]}\\
\times {}& \int_0^{\infty} \td{\rm SFR}_{10}\, p_z({\rm SFR}_{10} \mid M_\ast, {\rm SFR}_{100}) && \text{[burstiness]}\\
\times {}& \,p_z(M_{\rm UV} \mid {\rm SFR}_{10}, {\rm SFR}_{100}, M_\ast) ~. && \text{[UV mag.]}
\end{alignedat}
\label{eq:full_uvlf}
\end{equation}

Here, $p(M_{\ast}\mid M_{\rm h})$ represents the conditional probability of having a stellar mass $M_{\ast}$ in a halo of mass $M_{\rm h}$ (i.e., the stellar to halo mass relation, SHMR)\footnote{The upper limit of the integral over the conditional distribution of stellar masses is set by the physical limit of the average baryon mass in a halo of mass $M_{\rm h}$, $f_{\rm b}M_{\rm h}$, with $f_{\rm b}\equiv\Omega_{\rm b}/\Omega_{\rm m}$ the cosmic baryon fraction; we renormalize $p(M_\ast \mid M_{\rm h})$ accordingly.}. $p_z(\textrm{SFR}_{100}~|~M_\ast)$ defines the conditional probability of a star-formation rate averaged over $100$\,Myr given stellar mass, $M_\ast$ (i.e., the star-forming main sequence, SFMS). $p_z(\textrm{SFR}_{10}~|~M_\ast, {\rm SFR}_{100})$ represents the conditional probability of star formation rate averaged over shorter timescale, $10$\,Myr, at a fixed stellar mass and $100$\,Myr-averaged star formation (i.e., excess burstiness on shorter timescales). $p_z(M_{\rm UV} \mid {\rm SFR}_{10}, {\rm SFR}_{100}, M_\ast)$ describes the distribution of UV magnitudes conditioned on these quantities. The subscript $z$ indicates that these distributions may evolve with redshift, either through their means, variances, or both.  

Note that the choices of conditional distributions in Eq.~\eqref{eq:full_uvlf} are not unique.  Moreover, each term remains a conditional distribution on only a few variables. In reality, the stellar mass of a galaxy depends not only on its host halo mass, but also on other halo properties such as its assembly history and environment, and similarly for the other terms. We assume that marginalizing over these unmodeled dependencies yields tractable (log-normal) distributions whose scatter absorbs them, i.e., we treat them as subdominant rather than absent. For the clustering, this implies neglecting galaxy assembly bias, as the halo bias in our model depends on halo mass alone. This is a good approximation at high redshifts, where galaxies reside in rare halos far above the characteristic collapse mass, for which the dependence of halo clustering on assembly history is weak \citep{Gao2005, Wechsler2006, Wang2026, Yang2026}. We choose this basis because it offers a relatively direct and explicit connection between generic galaxy properties and the observables (UV LFs and ACFs) used in this work. One could alternatively directly parametrize the star formation histories of individual galaxies, choosing a parametric or non-parametric basis \citep[e.g.,][]{Carnall2019b, Leja2019,Endsley2024} and forward model the associated time evolution of the SEDs of individual galaxies using population synthesis models. However, in the absence of large, high S/N spectroscopic datasets with well-understood selection effects, inference using such a detailed basis could be biased and prior dependent \citep[e.g.,][]{Lower2020, Ji2022, Tacchella2022c,  Whitler2023, Haskell2024, Wan2024}.  As we detail below, the choice of basis in Eq.~\eqref{eq:full_uvlf} allows us to use well-motivated functional forms and associated priors, while still providing sufficient flexibility to characterize diverse galaxy models and observations.  In the following, we describe each component in more detail and motivate the adopted functional forms and priors.

\subsection{Halo mass function}
\label{sec:hmf}

Our galaxy model assumes that galaxies trace the underlying population of dark-matter halos. The number density of the latter is described by the HMF. Analytically, the HMF can be obtained from the excursion set formalism \citep{Press1974, Bond91}, with N-body simulations used to calibrate its exact shape \citep{Sheth2001, Reed2007, Lukic2007, Pillepich2010, Behroozi2013, Watson2013}.  In this work, we use the \citet{Tinker2008} HMF functional form calibrated to N-body simulations at $z<2.5$:\footnote{Halo mass functions have been computed using the {\tt hmf} package \citep{Murray2013}.}
\begin{equation}
    \frac{{\rm d}n}{{\rm d}M_{\rm h}} = f(\sigma)\frac{\bar{\rho}_{\rm m}}{M_{\rm h}}\frac{{\rm d}\ln\sigma^{-1}}{{\rm d}M_{\rm h}},
\end{equation}
with $f(\sigma) = A[\left( \frac{\sigma}{b}\right)^{-a}+1]{\rm e}^{-c/\sigma^2}$, where $\sigma$ is the rms of the linear density field smoothed on the scale $R(M_{\rm h})$, $\bar{\rho}_{\rm m}$ is the comoving mean matter density, and the parameters $A$, $a$ and $b$ evolve with redshift as in \citealt{Tinker2008} and are tuned to the simulation outputs. Halo masses are defined as $M_{200\rm m}$, the mass enclosed within 200 times the mean matter density.

The exact behavior of the HMF at higher redshifts is somewhat uncertain \citep[see][ for disagreement at the highest redshifts probed by JWST]{Yung_2024, OBrennan2024}. We have tested the impact of the choice of HMF on our results in Appendix~\ref{sec:appendixA}. We find that the posterior medians shift by at most $0.5\sigma$ between different HMFs.  We thus confirm that in the context of our model and data, the astrophysical uncertainties are far larger than uncertainties of the choice of HMF.\footnote{The choice of HMFs could impact results if the data were more constraining, or the theoretical model less flexible \citep[e.g.,][]{Mirocha2021, Greig2024}.  In those cases, it is important to either jointly sample parametric HMFs or use the Bayesian evidence to select between different HMFs \citep{Greig2024}.}

\subsection{Stellar-to-halo mass relation}
\label{sec:shmr}
A galaxy's stellar mass is linked to the mass of its dark matter halo through the stellar-to-halo-mass relation \citep[SHMR;][]{Behroozi2013, Harikane2016, Kravtsov2018, Shuntov2024}. Many simulations and observations find this relation to be approximately redshift-independent, with a log-normal scatter around the median \citep[e.g.,][]{More2011, Reddick2013, Wojtak2013, Zu2015, Ma2017, Ceverino2018, Girelli2020, Stefanon2021, Chaikin2025}. However, the normalization, slope, and scatter can vary dramatically between different models, at times resulting in orders of magnitude differences in the mean relation (see, e.g., Fig.~17 in \citealt{Yung2019} and Fig.~5 in \citealt{Davies2025}). 

Motivated by the above, we describe the SHMR using a log-normal conditional distribution for the stellar mass, $M_\ast$, given a host halo mass, $M_{\rm h}$: $p(\log M_\ast \mid \log M_{\rm h}) = \mathcal{N}(\log M_\ast ~|~ \mu_{M_{\ast}}(M_{\rm h}), \sigma_{\rm SHMR})$ \citepalias{Nikolic24}. The mean relation\footnote{We define $\mu_X$ as the mean of $\log X$ for every conditional relation in this work (e.g., $\mu_{M_\ast}$, $\mu_{\rm SFR_{100}}$), and $\sigma_X$ as the standard deviation of $\log X$, in dex; ``mean relation'' always refers to $\mu_X$. As $\log X$ is normally distributed, $10^{\mu_X}$ is the median of $X$, not its mean (see Appendix~B of \citetalias{Nikolic24}). Throughout, $\log\equiv\log_{10}$, and $\mathcal{N}(x \mid \mu, \sigma)$ is a normal distribution in $x$ with mean $\mu$ and standard deviation $\sigma$.} is given by

\begin{equation}
    \mu_{M_{\ast}} =
    \log{  \left( \min\left(f_{\ast,0}\,\frac{\left(\frac{10^{10}\,M_\odot}{M_{\rm knee}}\right) ^ {-\alpha_{\ast}} + \left(\frac{10^{10}\,M_\odot}{M_{\rm knee}}\right) ^ {0.61}}{\left(\frac{M_{\rm h}}{M_{\rm knee}}\right) ^ {-\alpha_{\ast}} + \left(\frac{M_{\rm h}}{M_{\rm knee}}\right) ^ {0.61}} M_{\rm h}, f_{\rm b} M_{\rm h} \right)\right)},
\label{eq:SHMR}
\end{equation}

\noindent where $f_{\ast,0}$ represents the normalization (the ratio $M_\ast/M_{\rm h}$ at $M_{\rm h}=10^{10}\,M_\odot$), $\alpha_{\ast}$ is the low-mass slope of $M_\ast/M_{\rm h}$, and $M_{\rm knee}$ is the characteristic mass where the slope changes and $10^{10}\,M_{\odot}$ is a reference halo mass scale \citep[e.g.,][]{Park2019}.  We enforce a physical upper limit on the stellar mass set by the cosmological baryon fraction (i.e., $M_\ast \leq f_{\rm b} M_{\rm h}$, with $f_{\rm b}\equiv\Omega_{\rm b}/\Omega_{\rm m}$ the cosmic baryon fraction).  A double power law is a common choice for the mean of the SHMR \citep[e.g.,][]{Shankar2006, Behroozi2010, Moster2010, Mirocha2017, Girelli2020, Shuntov2022}, with the fiducial explanation being that supernova feedback sets the low mass slope while feedback from active galactic nuclei sets the high mass slope (e.g., \citealt{Croton2006, Wyithe2013, Furlanetto2017, Wechsler2018}). As our results are not sensitive to the high-mass slope (the data probe halos well below $M_{\rm knee}$, to which they are also insensitive; Appendix~\ref{appendix:par_sweep}), we fix the high-mass exponent to $0.61$ (i.e., $M_\ast/M_{\rm h}\propto M_{\rm h}^{-0.61}$ for $M_{\rm h}\gg M_{\rm knee}$)  (\citealt{Behroozi2010, Mirocha2017}; \citetalias{Nikolic24}). We also confirm by running different inferences that our posteriors are effectively unchanged if we vary or fix $M_{\rm knee}$; hence in our default runs we fix its value to $2\times10^{12}\,M_{\odot}$, consistent with the halo mass at which the SHMR peaks in the observationally-calibrated abundance-matching literature \citep[e.g.,][]{Moster2010, Girelli2020, Shuntov2022}.  The scatter around the relation, $\sigma_{\rm SHMR}$ is treated as a free parameter.  For simplicity, we assume the scatter to be independent of mass and redshift, as found in some numerical simulations \citep[][but see \citealt{Matthee2017} and Sect.~\ref{sec:extensions}]{Lovell2021, Hassan2022}.  As we detail in Sect.~\ref{sec:extensions}, we use Bayesian evidence to confirm that the data do not show a strong preference for including a redshift or mass dependence to the SHMR.

\begin{figure}
    \centering
    \includegraphics[width=1.0\linewidth]{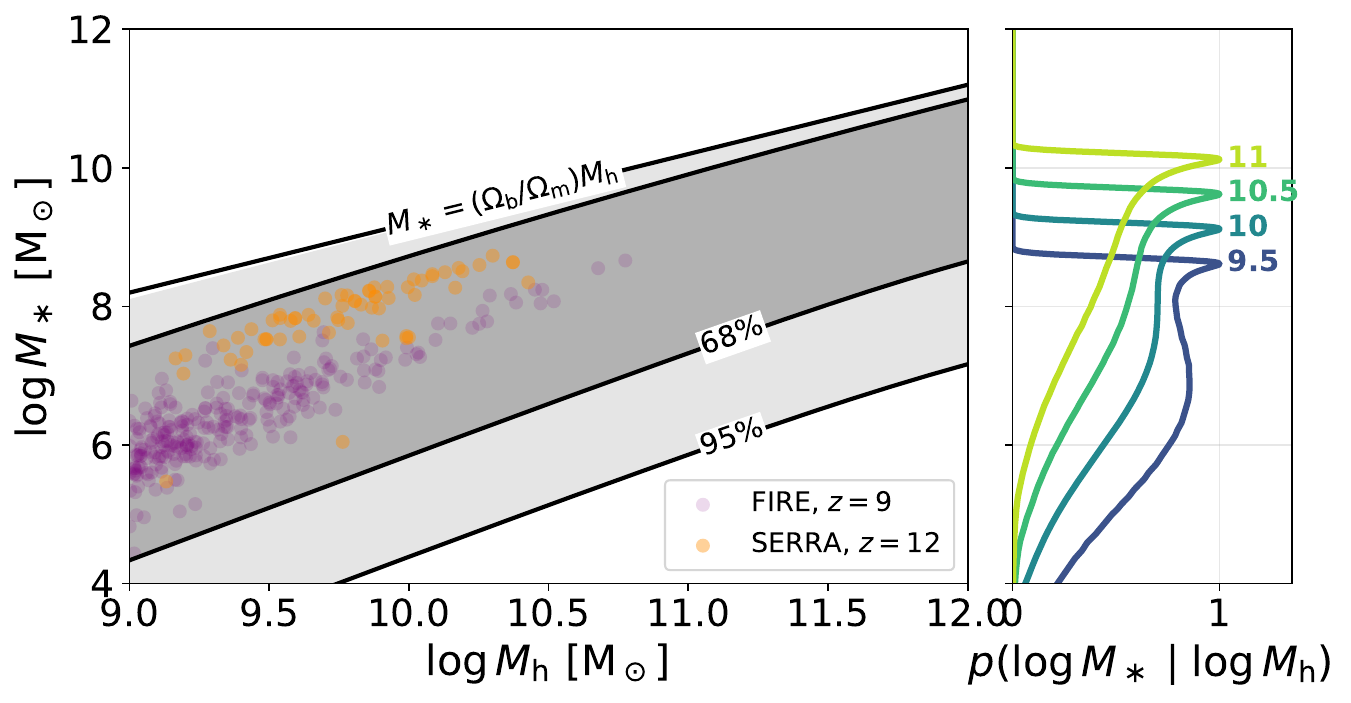}
    \caption{\textit{Left panel:} The joint prior distribution of stellar and halo masses in our fiducial model. The shaded regions show the 68\% and 95\% credible intervals of the prior, including both parameter uncertainty and intrinsic scatter (see text and Sect.~\ref{sec:priors}), and the black line the baryon-fraction limit $M_\ast=f_{\rm b}M_{\rm h}$. Points show galaxies from FIRE at $z=9$ \citep[purple;][]{Ma2017} and SERRA at $z=12$ \citep[orange;][]{Pallottini2022}. \textit{Right panel:} the corresponding prior over the conditional distributions of stellar mass at fixed halo mass (i.e., the SHMR), normalized to unit peak and labeled by $\log(M_{\rm h}/M_\odot)$. They peak just below $f_{\rm b}M_{\rm h}$ because, for part of the prior, the mean relation reaches this limit.}
    \label{fig:SHMR_prior}
\end{figure}

In summary, our fiducial functional form for the SHMR depends on 3 free parameters: ($f_{\ast,0}$, $\alpha_{\ast}$, $\sigma_{\rm SHMR}$).  The priors for these are determined by an ensemble of hydrodynamic simulations in Sect.~\ref{sec:priors}, and inferred from data in Sect.~\ref{sec:results}.

In Fig.~\ref{fig:SHMR_prior} we show the prior distribution for stellar and halo masses, together with results from the FIRE \citep{Ma2017} and SERRA \citep{Pallottini2022} simulations. The 68\% and 95\% credible intervals of the distributions in the left panel include the prior uncertainty in both the mean and the scatter around the mean; in the right panel we show the corresponding conditional PDFs (i.e., the SHMR).
The two hydrodynamic simulations predict different normalizations, but are contained (by construction; cf. Sect.~\ref{sec:priors}) in our prior.

\subsection{Star-formation history and the main sequence}

\label{sec:sfr}
\begin{figure}
    \centering
    \includegraphics[width=1.0\linewidth]{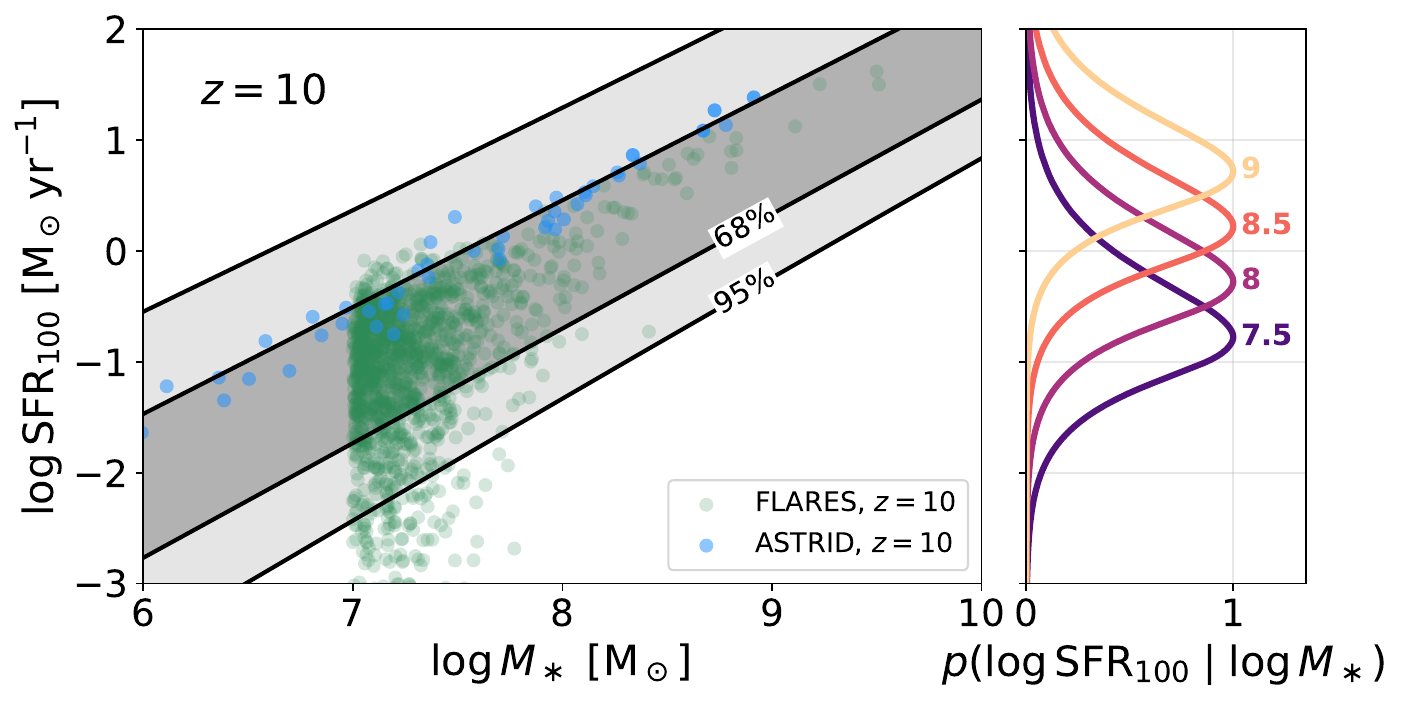}
    \caption{\textit{Left panel:} The joint prior distribution of SFR$_{100}$ and stellar masses at $z=10$ in our fiducial model. The shaded regions show the 68\% and 95\% credible intervals of the prior, including both parameter uncertainty and intrinsic scatter (see text). For context, the colored points show galaxies at $z=10$ from FLARES \citep[green;][]{Lovell2021} and ASTRID \citep[blue;][]{Bird2022}. \textit{Right panel:} the corresponding prior over the conditional distributions of $\mathrm{SFR}_{100}$ at fixed stellar mass (i.e., the SFMS), normalized to unit peak and labeled by $\log(M_\ast/M_\odot)$.}
    \label{fig:sfms_prior}
\end{figure}

The star-formation history (SFH) of a galaxy is one of its most fundamental properties.  How the total stellar mass discussed in the previous section is built-up over cosmic time remains unclear for most high-redshift galaxies. Because of relatively short cooling timescales at the virial radii of host halos, we expect the accretion of gas from the intergalactic medium (IGM) into the galaxy to be connected to the growth of the DM halo \citep{Keres2005, Dekel2009, Schaye2010}, which is on average exponential with time  (e.g., \citealt{McBride2009}).  If star formation is indeed limited by the accretion of gas from the IGM, such a scenario would naturally lead to an exponentially rising SFH for typical galaxies in the epoch of reionization (e.g., \citealt{Finlator2011, Behroozi2013, Endsley2024, Tang2026}). 
However, there is both observational and theoretical evidence to suggest large variations around such an ``average'' evolution (e.g., \citealt{Ma2017, Caplar2019, Faisst2019, Shen2023, Boyett2024, Endsley2024, Cole2025, Sun2025, NavarroCarrera2025}). Variations could be driven by clumpy accretion / mergers, radiative winds from massive stars, supernova feedback, etc., with each mechanism acting on a characteristic timescale (e.g., \citealt{Faucher-Giguere2018, Tacchella2020, Pallottini2023, Sun2026}). 

Characterizing such complex SFHs in order to interpret observations is very challenging.  Both parametric and non-parametric bases are routinely used to interpret JWST spectra (e.g., \citealt{Carnall2019b, Leja2019, Wan2024}; see \citealt{Caplar2019, Tacchella2020, Iyer2026} for power-spectral-density decompositions).

  These bases typically come with a large number of model choices and parameters (e.g., SFH bin sizes and associated covariances in non-parametric reconstruction, or the parametrization and priors of a power-spectral density decomposition), which can result in degeneracies, biases, and even unphysical solutions when interpreting galaxy spectra.  This has been directly demonstrated using observed spectra at $z\sim1$--3 \citep[e.g.,][]{Pacifici2015, Turner2025}, while at the highest redshifts ($z\gtrsim6$) similar biases have so far been shown primarily via photometric spectral energy distribution (SED) fitting \citep[e.g.,][]{Whitler2023}.  While some high S/N, broad coverage spectra can be constraining enough to distinguish between different SFH models \citep[e.g.,][]{Slob2024, Chen2026}, inferring detailed SFHs from most high-$z$ spectra still depends on the choice of basis and associated priors \citep[e.g.,][]{Pforr2012}. 
Here we take a less ambitious approach, focusing on the SFH averaged over the {\it timescales most directly probed by the observations we use in our likelihood}: photometric UV LFs and mass-complete ACFs.  The latter is sensitive to the total stellar mass, discussed in the previous section, while the former is sensitive to the star formation on timescales of $100$\,Myr \citep{Kennicutt1998, Kennicutt2012}.  We denote this as ${\rm SFR_{100}}$ and, unless explicitly stated otherwise, use it synonymously with the SFR throughout.  

Simulations and observations find a strong correlation between SFR$_{100}$ and $M_\ast$ \citep[e.g.,][]{Noeske2007, Daddi2007, Speagle2014, Ma2017, Katsianis2019, Popesso2023, Rinaldi2025}, commonly referred to as the star-forming main sequence, although simulations report SFRs averaged over different timescales (instantaneous for ASTRID and FLARES, 20\,Myr for SERRA, 100\,Myr for FIRE; see Sect.~\ref{sec:priors}). Motivated by these considerations, we model the conditional distribution between  SFR$_{100}$ and stellar mass with a log-normal: $p_z(\log {\rm SFR_{100}} \mid \log M_\ast) = \mathcal{N}(\log {\rm SFR_{100}} \mid \mu_{\rm SFR_{100}}(M_\ast, z), \sigma_{\rm SFMS}(M_\ast))$ with the mean given by
\begin{equation}
    \mu_{\rm SFR_{100}} (M_{\ast}, z)  = \log{\left( \frac{M_{\ast}}{t_{\ast} H^{-1}(z)}  \right)} \approx  \log{\left( \frac{M_{\ast}}{4.60\,t_{\ast} t_{\rm dyn}(z)}  \right)}, 
    \label{eq:sfms}
\end{equation}
where the free parameter $t_{\ast}$ represents the redshift-independent characteristic timescale for star-formation expressed as a fraction of the Hubble time, $H^{-1}(z)$, or analogously in terms of the halo dynamical time, $t_{\rm dyn}(z) \equiv \left(G\langle\rho\rangle_{\rm halo}\right)^{-1/2} = \sqrt{8\pi/(3\Delta_{\rm vir})}\,H^{-1}(z) \approx 0.217\,H^{-1}(z)$, evaluated at the mean halo overdensity predicted by the spherical collapse model during matter domination, $\langle \rho \rangle_{\rm halo} = \Delta_{\rm vir}\bar{\rho}(z) \approx 180\,\bar{\rho}(z)$ (with $\Delta_{\rm vir}=18\pi^2$). We use $t_{\rm dyn}$ when quoting star-formation timescales later in the paper (e.g., Eq.~\eqref{eq:sfms_map}). Such a functional form for the mean is expected from cosmological accretion-driven modes of galaxy growth \citep[e.g.,][]{Dekel2009, Bouche2010, Dave2012}.  Through $H(z)$, the mean SFMS in Eq.~\eqref{eq:sfms} evolves with redshift: at fixed $t_\ast$, the characteristic star-formation timescale decreases with increasing redshift, following the  dynamical time. We indicate this redshift dependence with the subscript $z$ in $p_z(\mathrm{SFR}_{100}\mid M_\ast)$.

Many simulations suggest that the scatter of the SFMS is mass-dependent (e.g., \citealt{Ceverino2018, Matthee2019, Katsianis2019, McClymont2025}), with larger variability found for lower mass systems. We model this using a linear relation at low stellar masses and a constant at high masses
\begin{equation}
    \sigma_{\rm SFMS} = \begin{dcases}
    a_{\sigma,\rm SFMS}\,\log\left(\frac{M_{\ast}}{10^{10}\,M_{\odot}}\right) + \sigma_{\rm SFMS,0}& \text{if } M_{\ast}\leq 10^{10}\,M_{\odot},\\
    \sigma_{\rm SFMS,0}              & \text{otherwise},
    \label{eq:sigma_sfms_equation}
\end{dcases}
\end{equation}
leaving both the slope ($a_{\sigma,\rm SFMS}$) and normalization ($\sigma_{\rm SFMS,0}$) as free parameters. At higher masses, the transition to a constant scatter can be justified by approaching a ``steady-state'' mode of star formation (e.g., \citealt{Ceverino2017, Matthee2019, Popesso2023}).

In the left panel of Fig.~\ref{fig:sfms_prior}  we show the 68\% and 95\% credible intervals of our prior in the space of the joint $p_z({\rm SFR}_{100},  M_\ast)$ distribution, obtained by varying the above mentioned mean and scatter over their respective prior ranges (discussed below).  For reference, we also show the values for galaxies from FLARES and ASTRID simulations \citep[][]{Lovell2021, Bird2022}.  Despite a notable difference in both the normalization of the mean relation and its scatter, the galaxy populations of both hydrodynamic simulations are well characterized by a log-normal distribution (see also \citealt{Matthee2019, Katsianis2019}).

In principle, our framework can be extended to connect different observables to star-formation on different timescales. Most notably, nebular emission lines (e.g., Balmer lines) in high-resolution spectra can inform the SFHs on shorter timescales \citep[$\sim10$\,Myr; e.g.,][]{CoveloPaz2026, Munoz2026, Tang2026, Gelli2026}. Doing this accurately, however, requires jointly modeling bespoke selection biases associated with each spectroscopic follow-up \citep[e.g.,][]{GagnonHartman2026, diCesare2026}, as well as evaluating the Bayesian evidence for different model and prior choices in order to see if the conclusions are physical or are instead being driven by the expanded (more complicated) models \citep[e.g.,][]{Trotta2008, Lawler2021}. 
We defer this to future work.

We do, however, account for star-formation variability on timescales shorter than 100\,Myr. Although the UV LFs are not very sensitive to it (see above), we include it for completeness. To compute the UV luminosity of a galaxy (Sect.~\ref{sec:uv_lum}), we input its SFH into the BPASS stellar population synthesis code \citep{Eldridge2017, Byrne2022}, which sums the spectra of the stellar populations formed at each age. We therefore construct, for each galaxy, an SFH consistent with its $M_\ast$ and SFR$_{100}$ that also includes this short-timescale variability. We model the SFR averaged over the last 10\,Myr, SFR$_{10}$, with the conditional distribution
\begin{equation}
\begin{split}
    p_z(\log{\rm SFR}_{10} &\mid \log{\rm SFR}_{100}) \\
    &= \mathcal{N}(\log{\rm SFR}_{10} \mid \log{\rm SFR}_{100},\, \sigma_{\rm SFR,10}) ~,
\end{split}
    \label{eq:sfr10}
\end{equation}
i.e., bursts and lulls symmetric (in log) around the 100\,Myr-averaged SFR\footnote{In principle, the mean of this relation could also be inferred, given a suitable prior. However, the data are not yet constraining, and we defer this to future work.}. The scatter $\sigma_{\rm SFR,10}$ is a free parameter of our inference. Between 10 and 100\,Myr before observation we take the SFR to be constant at SFR$_{100}$, so that the SFH averaged over the last 100\,Myr equals SFR$_{100}$ for $\sigma_{\rm SFR,10}\to0$ (and exceeds it by only $\approx1\%$ on average for $\sigma_{\rm SFR,10}=0.2$). More than 100\,Myr before observation, we assume a rising SFH, as motivated above, with the galaxy growing parallel to the mean SFMS, $\dot M_\ast = M_\ast H(z)/t_{\ast,i}$ with $t_{\ast,i}\equiv M_\ast/[{\rm SFR}_{100}H^{-1}(z)]$, which gives
\begin{equation}
    {\rm SFR}(z') = {\rm SFR}_{100}\,\frac{H(z')}{H(z_{100})}\left(\frac{1+z'}{1+z_{100}}\right)^{-1/t_{\ast,i}}, \qquad z'>z_{100},
    \label{eq:SFH_long}
\end{equation}
where $z_{100}$ is the redshift 100\,Myr before observation. We truncate the SFH at the lookback time at which its integral equals $M_\ast$, i.e., assuming no star formation at earlier times.

Each redshift is a separate population, so this SFH need not follow a galaxy through cosmic time. Instead, it gives BPASS a full star formation history that reproduces the SFR$_{10}$ and SFR$_{100}$ of each galaxy, so that the UV luminosity reflects star formation on both timescales. Our conclusions do not depend on its details.  Indeed, the UV LFs and mass-complete ACFs do not constrain $\sigma_{\rm SFR,10}$ (its posterior follows the prior), and within its prior it contributes only $\approx5\%$ of the variance in UV magnitude at $M_{\rm UV}=-20$ (Sect.~\ref{sec:res_stochasticity}).

\subsection{Metallicity}

The metal content of galaxies, quantified by the gas-phase metallicity ($Z$), can also impact their UV luminosities \citep[see, e.g., the  review in][]{Maiolino2019}. The gas-phase metallicity of galaxies is strongly correlated with the stellar mass as a result of cumulative metal production and retention, and to a lesser degree also the SFR. At low redshifts, metallicities are observed to follow the so-called fundamental metallicity relation (FMR; \citealt{Mannucci2010, Curti2020}), which can be represented as a 2D surface in the $M_{\ast}$–SFR–$Z$ space. 
JWST is now providing insight into the chemical evolution of galaxies out to high redshift ($z \gtrsim 8$). Early results indicate generally lower metallicities compared to the local Universe (for a given stellar mass), with a mass-metallicity relation already in place at $z \approx 8$ \citep[][]{Langeroodi2022, Curti2023, Kotiwale2025, Stanton2025}. However, the bursty nature of star formation which is likely more common at high redshift (see Sect.~\ref{sec:sfr}) also affects the scaling of metallicities, and the FMR in this population is still debated (\citealt{Heintz2023}; though see \citealt{Nakajima2023}). We leave more flexible treatment of metallicity scalings to future work, as current data are not yet sensitive to it, and these have a comparatively minor impact on the UV luminosities that we forward-model in this work.

We follow \citetalias{Nikolic24} and model the metallicity using a redshift-dependent FMR \citep{Curti2020, Curti2023}
\begin{equation}
    Z_\ast (M_\ast, {\rm SFR}_{100}, z) = 
    0.47 \left( 1+ \left( \frac{M_{\ast}}{M_0}\right)^{-2.1} \right)^{-0.148} 10^{\Delta_z} \ Z_{\odot} ~, 
    \label{eq:FMR}
\end{equation}
\noindent where $M_0(\mathrm{SFR}_{100}) \equiv 10^{10.11} \left( \frac{\mathrm{SFR}_{100}}{M_\odot\,\mathrm{yr}^{-1}} \right)^{0.56} M_\odot$. Following \citetalias{Nikolic24}, we convert the gas-phase FMR calibration of \citet{Curti2020,Curti2023} to a stellar metallicity via $Z/Z_\odot = 10^{(12+\log(\mathrm{O/H})-8.69)}$ \citep{Asplund2009}, adjusted for gas-phase abundances being higher than stellar metallicities by a factor of $\approx2.63$ on average \citep{Strom2018}; $Z_\ast (M_\ast, {\rm SFR}_{100}, z)$ above already includes this correction and represents the mean \emph{stellar} metallicity in units of solar metallicity, as required by the BPASS stellar population synthesis models used to compute UV luminosities (Sect.~\ref{sec:uv_lum}). The (tentative) redshift evolution of the relation is modeled as $\Delta_z = -0.056 z + 0.064$ from \citet{Curti2023}. We note that this parametrization is extrapolated to the highest redshifts probed by JWST, where observational constraints remain limited.  Furthermore, we do not include stochasticity in the metallicity (i.e., we ignore scatter around the above mean relation). \citetalias{Nikolic24} have shown that such scatter has a negligible impact on emissivities at high redshift (see also \citealt{Munoz2026}).

\subsection{UV magnitudes}
\label{sec:uv_lum}

We use the BPASS stellar population synthesis code \citep{Eldridge2017, Byrne2022} to compute the SED of a galaxy at $z$, given its SFH and metallicity, as described in the previous sections.  BPASS accounts for the presence of binary stars, which may become increasingly important at high redshift where lower metallicities and denser star-forming environments are expected to increase the binary fraction \citep{Badenes2018, Sana2025}.  This effectively results in a higher value of the $1500$\,\AA\ luminosity for a given  SFR$_{100}$, compared with single stellar populations (e.g., \citealt{Wilins2019, Mirocha2021, Yung2026})\footnote{For our fiducial model, we obtain a median value of  $\kappa_{\rm UV} \equiv {\rm SFR_{100}}/{L_{1500}}$ $\approx 6.9\times10^{-29}\,M_\odot\,{\rm yr^{-1}\,erg^{-1}\,s\,Hz}$, with roughly a $2\%$ scatter.}. 

We model the UV magnitude of a galaxy as a normal distribution.  Using the galaxy basis defined in the previous sections, we can write
\begin{equation}
\begin{split}
p_z (M_{\rm UV} \mid {}& \mathrm{SFR}_{10}, \mathrm{SFR}_{100}, M_\ast) = \\ & \mathcal{N}(M_{\rm UV} \mid \mu_{M_{\rm UV}}(\mathrm{SFR}_{10}, \mathrm{SFR}_{100}, M_\ast, z),\, \sigma_{\rm BPASS}) ~,
\label{eq:p_Muv}
\end{split}
\end{equation}
where $\mu_{M_{\rm UV}}$ is computed using BPASS\footnote{For simplicity we assign all stars of a galaxy a single metallicity, evaluated at the time of observation. Evaluating the FMR instead at the formation time of each stellar age bin (i.e., with the stellar mass, SFR$_{100}$ and redshift at that time) makes galaxies brighter by only $0.01$--$0.04$\,mag for $M_\ast=10^8$--$10^{10}\,M_\odot$ at $z=8$--$12$, as the UV luminosity is dominated by young stars and depends only weakly on metallicity.}, which we write schematically as
\begin{equation}
    \mu_{M_{\rm UV}} = {\rm BPASS} \left[{\rm SFH} ({\rm SFR}_{10}, {\rm SFR}_{100}, M_{\ast},z), Z_\ast\right] .
\end{equation}
In Eq.~\eqref{eq:p_Muv} we introduce an additional redshift- and magnitude-independent scatter, $\sigma_{\rm BPASS}=0.1$\,mag, that is not captured by the stochastic components of the model described above. This scatter may arise from scatter in the FMR \citep{Curti2020}, $\alpha$-element abundances \citep{Byrne2022} or IMF changes \citep{Stanway2019}.  Its fiducial value is consistent with the $<0.1$\,mag photometric color differences found across choices of stellar spectral library within BPASS by \citet{Byrne2023}. Note that assumptions about $\sigma_{\rm BPASS}$ have a negligible impact on our conclusions, as it is far subdominant compared to other sources of scatter in the model.

\section{Prior based on ensembles of hydrodynamic simulations}
\label{sec:priors}
\begin{figure}
    \centering
    \includegraphics[width=1.0\linewidth]{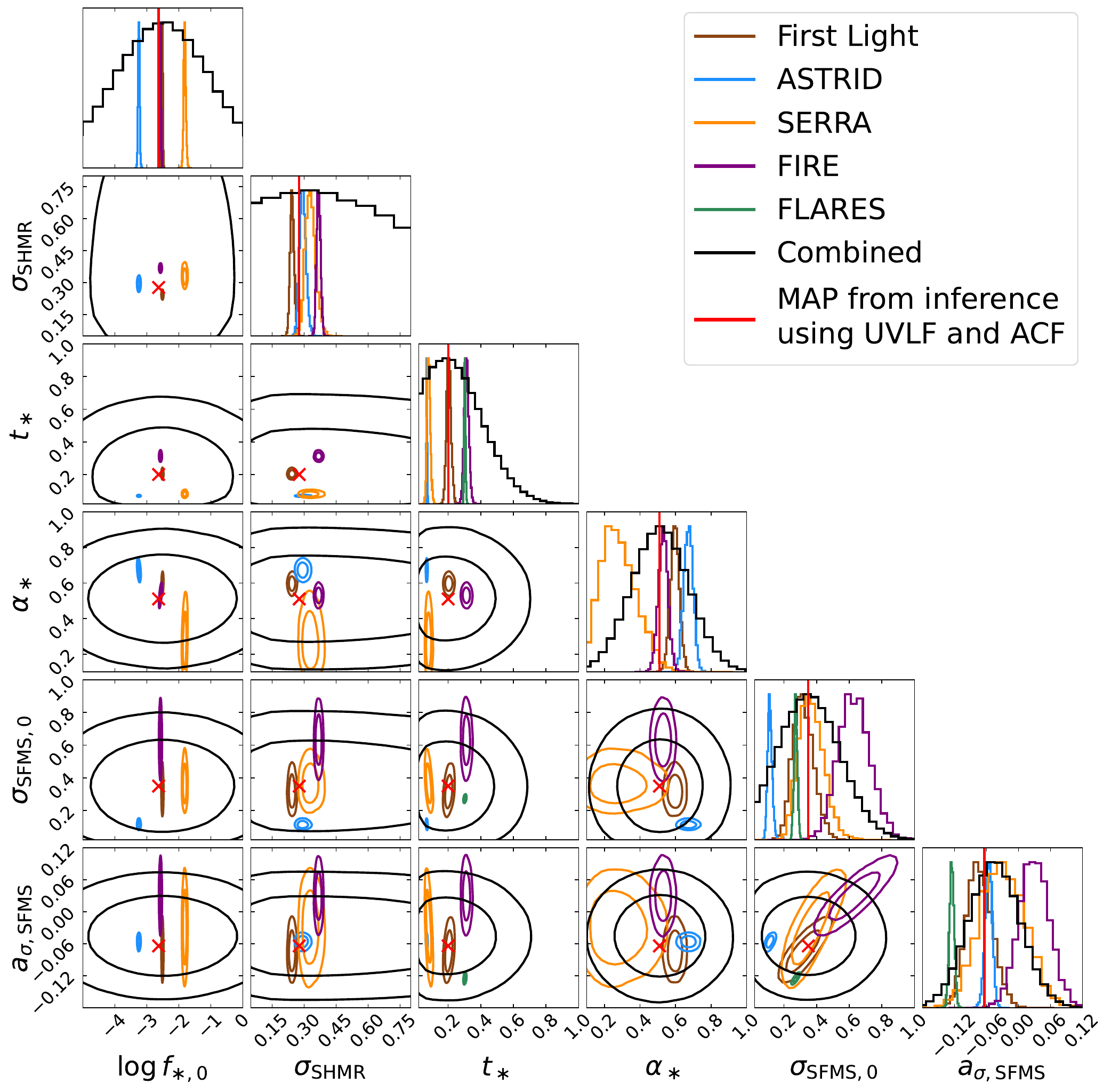}
    \caption{1D and 2D marginal corner plot showing the parameter constraints inferred from each simulation (colored contours, $68\%$ and $95\%$), i.e., ASTRID \citep{Bird2022}, FLARES \citep{Lovell2021}, First Light \citep{Ceverino2018}, SERRA \citep{Pallottini2022} and FIRE \citep{Ma2017}, along with the resulting combined prior (black contours, same levels); $\sigma_{\rm SFR,10}$, which the simulations do not constrain, is not shown.  We also indicate with the red crosses / vertical lines the MAP values inferred from JWST UV LFs and ACFs; the MAP is contained by our simulation-ensemble-based prior and yet is {\it inconsistent with any single simulation}.}
    \label{fig:prior_corner}
\end{figure}

As mentioned in Sect.~\ref{sec:galfor}, our galaxy formation model is rooted in conditional distributions, allowing for comparison with simulations that have these distributions as outputs. While different simulations adopt distinct subgrid prescriptions and are typically calibrated to reproduce low-redshift observables \citep[e.g.,][]{Crain2015}, they predict a range of scaling relations between galaxy and halo properties at high redshift. In order to compare the resulting distributions on an equal basis, we infer the posterior of the parameters for each simulation we consider, namely ASTRID \citep{Bird2022}, FLARES \citep{Lovell2021},\footnote{The public FLARES data do not include the halo masses of galaxies, so we use FLARES only to constrain the parameters of the SFMS.} First Light \citep{Ceverino2018}, SERRA \citep{Pallottini2022} and FIRE \citep[the FIRE-2 high-redshift suite;][]{Ma2017, Wetzel2023}. We treat every simulated galaxy as an independent draw from the log-normal SHMR and SFMS of Sect.~\ref{sec:galfor}, with uniform priors, using the snapshots closest to $z=5$, 10 and 15 for First Light, $z=5.5$, 10 and 12 for ASTRID (100 halos per snapshot, drawn evenly in logarithmic halo mass), $z=6$ and 12 for SERRA, $z=5$, 7 and 9 for FIRE, and $z=5$ and 10 for FLARES. We take the simulated SFRs as provided, i.e., averaged over different timescales (Sect.~\ref{sec:sfr}).  For FIRE, whose public catalogs do not include SFRs, we compute them from the formation times of the star particles, averaged over the last $100$\,Myr.\footnote{This choice mainly affects the SFMS scatter: for FIRE, using 10 instead of 100\,Myr changes $t_\ast$ by only $\approx10\%$ (0.28 vs.\ 0.31), but increases $\sigma_{\rm SFMS,0}$ from 0.63 to 0.93.}

The resulting parameter constraints are shown in Fig.~\ref{fig:prior_corner}, with the posterior for each simulation represented by a distinct color. The posteriors are relatively narrow, but they are disjoint in the joint parameter space, with distances between them much larger than their individual widths, even though some one-dimensional marginals overlap. Once expressed in this common space, it is evident that these simulations predict very different high-$z$ galaxy populations.  This is evident when comparing them in the relatively high-dimensional space of Fig.~\ref{fig:prior_corner}; if instead we compared them in a lower-dimensional space of, for example, the abundances of UV bright galaxies at a few redshifts, some would seem to predict similar galaxy populations \citep[e.g.,][]{Ceverino2017, Ma2017, Vijayan2021, Bird2022}.  This highlights the importance of comparing models in a ``universal'' space of sufficiently-high dimensionality.

We construct our prior to encompass all of the simulation results. To do so, we pool the posterior samples of all simulations into a single set and adopt a Gaussian prior with the same mean and variance, for each parameter, as this pooled set. The prior width therefore reflects both the uncertainty of the individual fits and, mainly, the differences between the simulations. We treat all simulations equally (each contributes $\approx6000$--$7000$ samples to the pool; FLARES, which constrains only the SFMS, contributes only to the SFMS parameters) and do not impose additional weighting, given the lack of a clear criterion for favoring one model over another. We do not keep the correlations between parameters in the pooled set, as they mainly reflect where the individual simulations lie in parameter space rather than physical relations between the parameters.

The spread between simulations is, however, not a complete measure of our theoretical uncertainty, and for three parameters we adopt wider prior widths than the pooled posteriors imply. First, all simulations predict a similar scatter in the SHMR, $\sigma_{\rm SHMR}\approx0.24$--$0.37$ (pooled standard deviation $0.05$). This agreement between a handful of simulations with comparable feedback implementations and halo selections is not independent evidence for a particular value, and adopting it as a prior would fix one of the main sources of stochasticity we aim to constrain. We therefore adopt a width of $0.6$ for $\sigma_{\rm SHMR}$, i.e., an effectively uninformative prior over its physically allowed range. Because the normalization of the SHMR is degenerate with its scatter in the UV LF (Sect.~\ref{sec:observables}), we correspondingly widen the prior on $\log f_{\ast,0}$ to $1.55$\,dex, allowing it to follow this degeneracy. Second, the simulations provide SFRs averaged over different timescales, from instantaneous rates (ASTRID, FLARES) to $100$\,Myr averages (FIRE), whereas our SFMS describes the SFR averaged over $100$\,Myr. They also fall into two distinct groups in $t_\ast$ ($\approx0.07$--$0.08$ for ASTRID and SERRA, $\approx0.2$--$0.3$ for First Light, FLARES and FIRE), and they derive SFRs and stellar masses with different definitions (fixed apertures, fractions of the virial radius or halo-finder membership). As the effect of these choices on $t_\ast$ cannot be quantified from the public data alone, we conservatively double the pooled width of $t_\ast$ to $0.21$; using the pooled width instead shifts the posterior median of $t_\ast$ by $0.4\sigma$ and all other parameters by less than $0.1\sigma$. Since shorter averaging timescales increase the scatter around the SFMS, the simulation values of $\sigma_{\rm SFMS,0}$ are, if anything, upper limits for our SFMS; for this and all remaining parameters, the prior width is the pooled spread of the simulations.

Finally, we truncate the prior to physically allowed ranges, most importantly requiring the scatters ($\sigma_{\rm SHMR}$, $\sigma_{\rm SFMS,0}$) and the star-formation timescale ($t_\ast$) to be positive. The simulations do not constrain the burstiness on 10\,Myr timescales, and for $\sigma_{\rm SFR,10}$ we adopt a Gaussian prior with mean $0.2$ and standard deviation $0.045$, motivated by observations: the Balmer-line (tracing $\approx10$\,Myr) to UV (tracing $\approx100$\,Myr) luminosity ratio scatters by $0.3$--$0.46$\,dex at $z\approx1.4$--7, and the SFMS is correspondingly broader when measured with Balmer-line SFRs, by $\approx0.4$\,dex added in quadrature \citep{Clarke2024, Cole2025}. As these values also include variations in dust attenuation and in the conversion to SFR, they are upper limits on $\sigma_{\rm SFR,10}$, and our prior lies below them. 

Black contours in Fig.~\ref{fig:prior_corner} show the 68\% and 95\% credible regions of our constructed prior, and Table~\ref{tab:params} lists the prior of each parameter. The resulting prior encompasses the full range of parameter values predicted by the simulations, but excludes unphysical values. It therefore encodes our current theoretical uncertainty in galaxy formation at high redshift, as informed by state-of-the-art simulations.

For reference in Fig.~\ref{fig:prior_corner} we also indicate with red crosses the parameter values corresponding to the maximum a posteriori (MAP) model from our inference of UV LF and ACF observations.  Interestingly, while the MAP model lies within our simulation-ensemble-based prior, there is no single simulation that matches it.  In other words we find that none of the galaxy populations predicted by the simulations is consistent with the one implied by the observations.

\section{From the model to the observables}
\label{sec:observables}
We use our galaxy model described in Sect.~\ref{sec:galfor} to compute observables that are compared to existing data using Bayesian inference. In this section we detail how we compute UV luminosity functions (Sect.~\ref{sec:uvlf}) and mass-complete angular correlation functions (Sect.~\ref{sec:acf}), as well as the observational datasets they are compared with.

\subsection{UV luminosity function}
\label{sec:uvlf}

We compute the UV luminosity function following Eq.~\eqref{eq:full_uvlf}, which integrates over the HMF and the hierarchy of conditional distributions summarized in Table~\ref{tab:model}.

\begin{table*}
\caption{Components of the UV LF model (Eq.~\eqref{eq:full_uvlf}). We use the \citet{Tinker2008} HMF. All conditional distributions are log-normal (normal in $M_{\rm UV}$, with a fixed $\sigma_{\rm BPASS}=0.1$\,mag); their mean and scatter are given by the listed equations.}
\label{tab:model}
\centering
\small
\begin{tabular}{@{}llll@{}}
\hline\hline
Component & Distribution & Free parameters & Eq. \\
\hline
Halo abundance  & ${\rm d}n/{\rm d}M_{\rm h}$ & -- & Sect.~\ref{sec:hmf} \\
SHMR            & $p(M_\ast \mid M_{\rm h})$ & $f_{\ast,0}$, $\alpha_\ast$, $\sigma_{\rm SHMR}$ & \ref{eq:SHMR} \\
SFMS            & $p_z({\rm SFR}_{100} \mid M_\ast)$ & $t_\ast$, $\sigma_{\rm SFMS,0}$, $a_{\sigma,\rm SFMS}$ & \ref{eq:sfms}, \ref{eq:sigma_sfms_equation} \\
Burstiness      & $p_z({\rm SFR}_{10} \mid {\rm SFR}_{100})$ & $\sigma_{\rm SFR,10}$ & \ref{eq:sfr10} \\
UV magnitude    & $p_z(M_{\rm UV} \mid {\rm SFR}_{10}, {\rm SFR}_{100}, M_\ast)$ & -- & \ref{eq:p_Muv} \\
\hline
\end{tabular}
\end{table*}

Each conditional distribution encodes one component of the galaxy formation model described in Sect.~\ref{sec:galfor}, and has a clear interpretation, allowing us to use well-motivated functional forms and corresponding priors (see Sect.~\ref{sec:priors}). For a given set of galaxy parameters, we numerically evaluate Eq.~\eqref{eq:full_uvlf}.  We suppress star formation in the smallest halos by multiplying the HMF by $\exp(-M_{\rm turn}/M_{\rm h})$, with $M_{\rm turn}=5\times10^8\,M_\odot$ \citep{Park2019}. This mimics the inefficient cooling of gas in halos close to the atomic cooling threshold ($T_{\rm vir}\approx10^4$\,K, i.e., $M_{\rm h}\sim10^8\,M_\odot$ at $z\approx10$; \citealt{Barkana2001, Oh2002}), as well as photoheating \citep{Sobacchi2013} and supernova feedback, which suppress star formation in halos up to a few $\times10^8$--$10^9\,M_\odot$.\footnote{In the faintest fitted bins, half of the galaxies reside in halos with $M_{\rm h}\lesssim10^{9.6}$--$10^{9.9}\,M_\odot$. Changing $M_{\rm turn}$ to $10^8$ ($10^9$)$\,M_\odot$ changes the model UV LF there by $\approx+8$--$9\%$ ($-8$--$9\%$) at $z=8$--$10$ and $+15\%$ ($-14\%$) at $z=12$, well below the measurement uncertainties.} 

We also include a simple dust treatment, which serves to dim the brightest galaxies. We follow the $A_{\rm UV} - \beta$ parametrization of \citet{Meurer1999} \citep[see also][]{Kar2026}. For $z<8$, we adopt the $\beta$--$M_{\rm UV}$ relation from \citet{Bouwens2014}. For $8 \leq z \leq 10$, we follow the JWST-derived $\beta$--$M_{\rm UV}$ relation from \citet{Cullen2024}. At higher redshifts we assume $\beta=-2.5$, for which the \citet{Meurer1999} relation gives $A_{\rm UV}=0$.  Dust is not applied in the likelihood: to avoid extending our model to include a more complete dust treatment with additional free parameters, we restrict our analysis of UV LFs to relatively faint galaxies, $M_{\rm UV}>-20$, where $A_{\rm UV}\lesssim0.5$\,mag in the brightest fitted bin and $A_{\rm UV}=0$ for $M_{\rm UV}\gtrsim-18.7$.  The dust model therefore serves only to better visualize the UV LFs over a broader range of magnitudes and does not impact our conclusions.  In future work we will extend our galaxy model to include dust via additional conditional distributions (e.g., \citealt{Sommovigo2026}), allowing us to extend our interpretation also to the bright end of the UV LFs.

Several groups have estimated the UV LFs from JWST data, using different fields, analysis pipelines and redshift bins. These estimates disagree with each other, especially at the highest redshifts (Fig.~\ref{fig:uvlf_post}), and fitting them together would require a model of their systematic differences (e.g., cosmic variance between fields, selection and completeness corrections), which is beyond the scope of this work. We therefore fit the UV LFs of a single group at all redshifts, measured from the same fields with the same pipeline, and show in Appendix~\ref{sec:app_inf_uvlf_others} how the results change when the UV LFs of other groups \citep{Donnan2024, Harikane2025, Whitler2025} are used instead.  As our fiducial observation, we use the $M_{\rm UV}>-20$ UV LFs at $z=8$--$12$ from \citet{Willott2024}. \citet{Willott2024} estimates the UV LF from the CANUCS survey targeting 10 fields across 5 independent sightlines towards strong-lensing galaxy clusters. Lensing enables constraints down to fainter magnitudes ($M_{\rm UV} \approx -18$), which are not sensitive to our dust model as discussed above.  Furthermore, the strategy of measuring the UV LF using $5$ independent sightlines minimizes potential biases due to cosmic variance (e.g., \citealt{Trapp2023}). 

The choice of UV LF observations used in the likelihood can make a notable impact on the inferred galaxy properties. Our fiducial choice of the \citet{Willott2024} estimate yields SHMR posteriors consistent with those from the spectroscopically confirmed UV LF of \citet{Harikane2025}.

\subsection{Angular correlation functions}
\label{sec:acf}

In addition to UV LFs, we use JWST observational estimates of ACFs to inform our galaxy model.  We forward-model the ACFs with the halo model \citep{Seljak2000, Peacock2000, Cooray2002}, populating halos with galaxies through a halo occupation distribution (HOD; \citealt{Berlind2002, Zheng2005}) derived from our SHMR, as implemented in the \texttt{halomod} package \citep{Murray2021}, with the halo bias of \citet{Tinker2010}.  The halo model describes the clustering of galaxies as the sum of one- and two-halo terms, corresponding to pairs of galaxies within the same halo and in different halos, respectively.  We fit the ACFs only on scales $\theta>0.005^\circ$ ($\approx0.8$\,cMpc at $z\approx7$--$9$), which are dominated by the two-halo term \citep[e.g.,][]{Jose2016, Bhowmick2018}. Galaxies above the mass-completeness limit of the clustering samples reside in rare, massive halos and are predominantly centrals \citep{Harikane2016, Harikane2018, Bhowmick2018}. The small-scale signal is therefore more susceptible to interlopers and systematics, and fitting it would make our results sensitive to the assumed satellite and interloper populations.

To compute the 3D galaxy correlation function, $\xi$, from which we obtain the ACF below, {\tt halomod} uses the HOD framework to assign galaxies to halos.  Specifically, a central galaxy sits at the center of its host halo, while satellite galaxies are hosted by sub-halos surrounding the central. We start by writing the mean relation for the occupation of central galaxies (which follows a Bernoulli distribution; e.g., \citealt{Zheng2005})
\begin{equation}
    N_{\rm c}(M_{\rm h}) = \frac{\Phi(x_{\rm b}) - \Phi(x_{\rm lim})}{\Phi(x_{\rm b})},
    \label{eq:centrals}
\end{equation}
with
\begin{equation}
    x_{\rm b} = \frac{\log(f_{\rm b} M_{\rm h}) - \mu_{M_\ast}(M_{\rm h})}{\sigma_{\rm SHMR}}, \quad
    x_{\rm lim} = \frac{\log M_{\ast,\rm lim} - \mu_{M_\ast}(M_{\rm h})}{\sigma_{\rm SHMR}},
    \label{eq:Phi_def}
\end{equation}
where $\Phi(x) = \frac{1}{2}\,\textrm{erfc}(-x/\sqrt{2})$ is the cumulative distribution function of the standard normal distribution.  In the above $\mu_{M_\ast}(M_{\rm h})$ is the mean SHMR from Eq.~\eqref{eq:SHMR}, $M_{\ast,\rm lim}$ is the mass-completeness limit of the clustering sample, and we fix $N_{\rm c}(M_{\rm h}) \equiv 0$ whenever $f_{\rm b}M_{\rm h} < M_{\ast,\rm lim}$, i.e., below the halo mass at which the baryon-fraction ceiling drops beneath $M_{\ast,\rm lim}$. This avoids unphysical parameter combinations where the stellar mass is larger than the baryon mass (similarly to Eq.~\eqref{eq:full_uvlf}).  In the limit $\log(f_{\rm b}M_{\rm h}) - \mu_{M_\ast}(M_{\rm h}) \gg \sigma_{\rm SHMR}$, the baryon-fraction ceiling lies far in the upper tail of the distribution, $\Phi(x_{\rm b}) \to 1$, and Eq.~\eqref{eq:centrals} reduces to the familiar non-truncated form $N_{\rm c}(M_{\rm h}) \to 1-\Phi(x_{\rm lim})$. 
Equation~\eqref{eq:centrals} is the probability that the central galaxy of a halo of mass $M_{\rm h}$ lies above $M_{\ast,\rm lim}$, given the same log-normal SHMR scatter that enters the UV LF (Eq.~\eqref{eq:SHMR}), truncated at $f_{\rm b}M_{\rm h}$. Halos somewhat below the mass at which the mean SHMR reaches $M_{\ast,\rm lim}$ therefore contribute to the clustering sample if their central scatters up in $M_\ast$, and halos somewhat above it are partly missed. Thus, $M_{\ast,\rm lim}$ sets the halo mass around which galaxies enter the sample, and $\sigma_{\rm SHMR}$ sets how gradual this transition is. Because lower-mass halos are more numerous, a larger scatter brings more of them into the sample and lowers its clustering amplitude. 
As our fit is restricted to scales dominated by the two-halo term, we fix the mean number of satellites to $N_{\rm s}(M_{\rm h}) = N_{\rm c}(M_{\rm h})\,[(M_{\rm h} - M_{\rm cut})/M_1]^{\alpha_{\rm sat}}$ for $M_{\rm h}>M_{\rm cut}$ (and zero otherwise; $N_{\rm s}$ is the mean number of satellites above $M_{\ast,\rm lim}$, e.g., $N_{\rm s}\approx3$ for $M_{\rm h}=10^{13}\,M_\odot$), with $M_{\rm cut}=10^{11.65}\,h^{-1}M_\odot$, $M_1=10^{12.3}\,h^{-1}M_\odot$ and $\alpha_{\rm sat} = 1$ based on the inferences of \citet{Paquereau2025} and \citet{Shuntov2025} and the simulations of \citet{Bhowmick2018}. With these values, the one-halo term contributes $<0.1\%$ of the ACF on the fitted scales (compared to $\approx15\%$ at $\theta=0.001^\circ$), and shifting $M_{\rm cut}$ or $M_1$ by $\pm0.5$\,dex changes the model ACF there by $<0.5\%$ \citep[see also Appendix of][]{Shuntov2025}.

The ACF in the halo model can then be expressed as
\begin{equation}
    \omega(\theta) = \int \td z_1 \int \td z_2 \phi(z_1) \phi(z_2) \xi \left( r(z_1,z_2,\theta)\right),
\end{equation}
where $\phi(z)$ is the normalized redshift distribution of the sample \citep{Paquereau2025} and $\xi$ is the 3D galaxy correlation function, including the one- and two-halo terms of both central and satellite galaxies.  An ACF estimated from a finite field is biased low by the integral constraint \citep{Groth1977}, the mean of $\omega$ over all pairs of positions in the field,
\begin{equation}
    \omega_{\rm IC} = \frac{1}{\Omega_{\rm s}^2} \int \int \omega(\theta_{12})\, \td\Omega_1 \td\Omega_2,
\end{equation}
where $\Omega_{\rm s}$ is the solid angle of the field and $\theta_{12}$ the separation between the solid-angle elements $\td\Omega_1$ and $\td\Omega_2$. We approximate the COSMOS-Web field \citep{Casey2023} as a $41.5'\times46.6'$ rectangle ($0.54$\,deg$^2$) and compare $\omega(\theta)-\omega_{\rm IC}$ to the data.
\begin{figure}
    \centering
    \includegraphics[width=\linewidth]{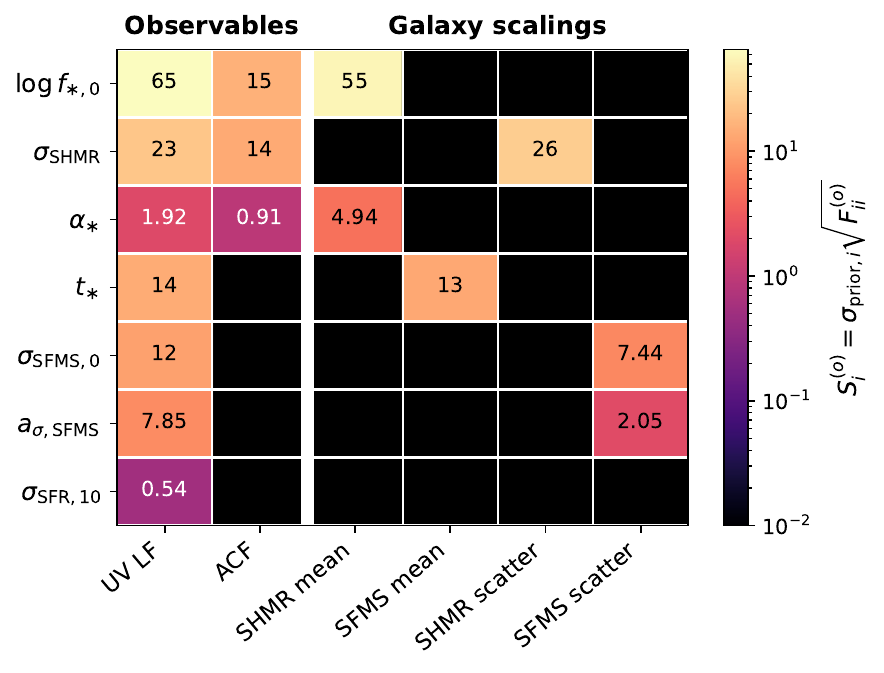}
    \caption{Sensitivity of each observable and scaling relation (columns) to each free parameter (rows), for hypothetical measurements with a $20\%$ uncertainty in each of 8 bins: the UV LF at $z=7.99$ ($-21\leq M_{\rm UV}\leq-17.5$), the ACF at $z=7.29$ for $M_{\ast,\rm lim}=10^9\,M_\odot$ ($0.005^\circ\leq\theta\leq0.05^\circ$), the mean and scatter of the SHMR ($10^{9.5}\leq M_{\rm h}/M_\odot\leq10^{11.5}$), and the mean and scatter of the SFMS at $z=7.99$ ($10^{8}\leq M_\ast/M_\odot\leq10^{10}$). Each cell shows $S_i=\sigma_{{\rm prior},i}\sqrt{F^{(o)}_{ii}}$, the change of observable $o$, in units of its uncertainty, produced by a $1\sigma$ prior change of parameter $\lambda_i$ around the fiducial value (here set to the median of the marginal posterior from Sect.~\ref{sec:results}); $S_i>1$ means that the observable is informative about $\lambda_i$ beyond the prior. Black cells indicate that the quantity does not depend on the parameter.}
    \label{fig:sensitivity}
\end{figure}

We compare the forward-modeled ACFs to the observational estimates of \citet{Paquereau2025}. They estimate the angular correlation function for galaxies from the COSMOS-Web survey using a mass-limited sample out to $z=10.5$. Using angular correlation with a mass-limited sample allows us to directly connect the results to our model, and most directly constrain the connection between stellar mass and halo mass, without the need to account for completeness in traditional emission-line limited samples. As our fiducial we use the results from redshift bins $z\in[6,8]$ and $z\in[8,10.5]$. We use $M_{\ast,\rm lim}=10^9\,M_\odot$ as the limiting stellar mass. This is the lowest limit for which our model fits the clustering well: for the $10^{8.7}\,M_\odot$ sample the fit is poor ($\chi^2\approx23$ for 10 data points), possibly due to incompleteness, while the $10^{9.3}\,M_\odot$ sample has larger uncertainties. With the poorly fit $10^{8.7}\,M_\odot$ or the noisier $10^{9.3}\,M_\odot$ sample, only the SHMR scatter changes noticeably, moving up to values similar to those from the UV LFs alone (Sect.~\ref{sec:res_shmr}).

\subsection{Sensitivity of the observables to the model parameters}
\label{sec:sensitivity}

To anticipate which parameters each observable can constrain, we quantify how strongly each observable responds to them using a Fisher-matrix approach \citep{Tegmark1997}. For each ``observable'' $o$ evaluated in bins $k$, we define the data vector $d_k=\log y_k$, where $y_k$ is the model prediction in bin $k$, and we approximate the Fisher matrix as\footnote{In reality, the covariance between bins is not diagonal (e.g., \citealt{Huang2026}), and the second term that corresponds to the change in the covariance might not be negligible (e.g., \citealt{Prelogovic2023}). We defer a more detailed exploration of the shape of the likelihood to future work.}
\begin{equation}
\label{eq:Fisher}
F^{(o)}_{ij} = \sum_{k \in o} \frac{1}{\sigma_k^2}
\frac{\partial d_k}{\partial \lambda_i}
\frac{\partial d_k}{\partial \lambda_j} ~.
\end{equation}
We compute the predictions with the same forward model used in the inference, and evaluate the derivatives at the posterior median of each 1D marginal from our fiducial inference, presented in Sect.~\ref{sec:results} (Table~\ref{tab:params}) with symmetric finite differences of $\pm0.1$ posterior standard deviations. Here $\lambda_i$ are the free parameters of the model and $\sigma_k$ is the uncertainty of $d_k$. To compare the observables on an equal footing, we assume a $20\%$ uncertainty in each of 8 bins of every quantity, i.e., $\sigma_k=\log1.2\approx0.08$\,dex. As the diagonal elements depend on the units of the parameters, we show in Fig.~\ref{fig:sensitivity}
\begin{equation}
S^{(o)}_i = \sigma_{{\rm prior},i}\sqrt{F^{(o)}_{ii}},
\end{equation}
the change of observable $o$, in units of its uncertainty, produced by a $1\sigma$ change of $\lambda_i$ within its prior (Sect.~\ref{sec:priors}); $S^{(o)}_i>1$ means that $o$ is informative about $\lambda_i$ beyond the prior. In Appendix~\ref{appendix:par_sweep} we show the results of the same analysis where instead of a constant $20\%$ uncertainty we use the actual uncertainties of the data.

The ACF depends only on the SHMR parameters, as the clustering sample is selected using stellar mass (Eq.~\eqref{eq:centrals}). It is sensitive mainly to $f_{\ast,0}$ and $\sigma_{\rm SHMR}$ ($S\approx15$ and $14$), which set the halos that host galaxies above $M_{\ast,\rm lim}$, and only weakly to $\alpha_\ast$ ($S\approx0.9$), because these galaxies occupy a narrow range of halo masses. The UV LF responds to all parameters: most strongly to $f_{\ast,0}$ ($S\approx65$), then to $\sigma_{\rm SHMR}$, $t_\ast$ and $\sigma_{\rm SFMS,0}$ ($S\approx12$--$23$), to $a_{\sigma,\rm SFMS}$ ($S\approx8$) and to $\alpha_\ast$ ($S\approx1.9$). At equal precision, the ACF is thus about $60\%$ as sensitive to $\sigma_{\rm SHMR}$ as the UV LF ($S=14$ vs.\ $23$), but four times less sensitive to $f_{\ast,0}$.

The value of the ACF lies in the different combination of parameters it constrains. Increasing $f_{\ast,0}$ or $\sigma_{\rm SHMR}$ by $0.1$ prior standard deviations raises the UV LF at $M_{\rm UV}=-19$ ($z\approx8$) by $41\%$ and $13\%$, respectively, while it lowers the ACF by $\approx10\%$ for either (not shown): a higher $f_{\ast,0}$ shifts galaxies above $M_{\ast,\rm lim}$ into less massive halos, whereas a larger $\sigma_{\rm SHMR}$ mixes in galaxies up-scattered from the far more abundant low-mass halos (Sect.~\ref{sec:acf}). In both observables an increase of one parameter can therefore be compensated by a decrease of the other, but the ACF weights $\sigma_{\rm SHMR}$ about three times more strongly relative to $f_{\ast,0}$, and it is insensitive to the SFMS parameters, which are degenerate with both in the UV LF. Adding the ACF should therefore narrow the posterior along the $f_{\ast,0}$--$\sigma_{\rm SHMR}$ degeneracy, mainly in $\sigma_{\rm SHMR}$ (Sect.~\ref{sec:res_shmr}).

\section{Results}
\label{sec:results}
We infer the model parameters from the datasets described in Sect.~\ref{sec:observables}, using the simulation-based priors of Sect.~\ref{sec:priors}.
We sample the posterior with \textsc{MultiNest} \citep{Fehroz2009}. We assume a Gaussian likelihood for both observables, using the reported observational uncertainties\footnote{We only have access to the diagonal variances.  By ignoring off-diagonal elements of the covariance, we expect our posteriors to be somewhat overconfident \citep[e.g.,][]{Prelogovic2023, Huang2026}. In future work we plan to use simulation based inference in order to avoid having to specify a functional form for the UV LF and ACF likelihoods entirely (e.g., Gagnon-Hartman et al. in prep.).}. For the UV LFs, the Gaussian is in the number densities themselves, not in their logarithm; as the reported uncertainties are asymmetric, we use the form of \citet{Barlow2004}, in which the width of the Gaussian varies linearly between the lower and upper uncertainty. For the ACFs, we treat the angular bins as independent. Throughout, we compare two inferences: one using only the UV LFs in the likelihood (``UVLF only'') and one using both the UV LFs and the ACFs (``UVLF + ACF''). We first show that the model reproduces both datasets and then turn to the model parameters; the full posterior is given in Appendix~\ref{appendixB}.

\begin{figure}
    \centering
    \includegraphics[width=1.0\linewidth]{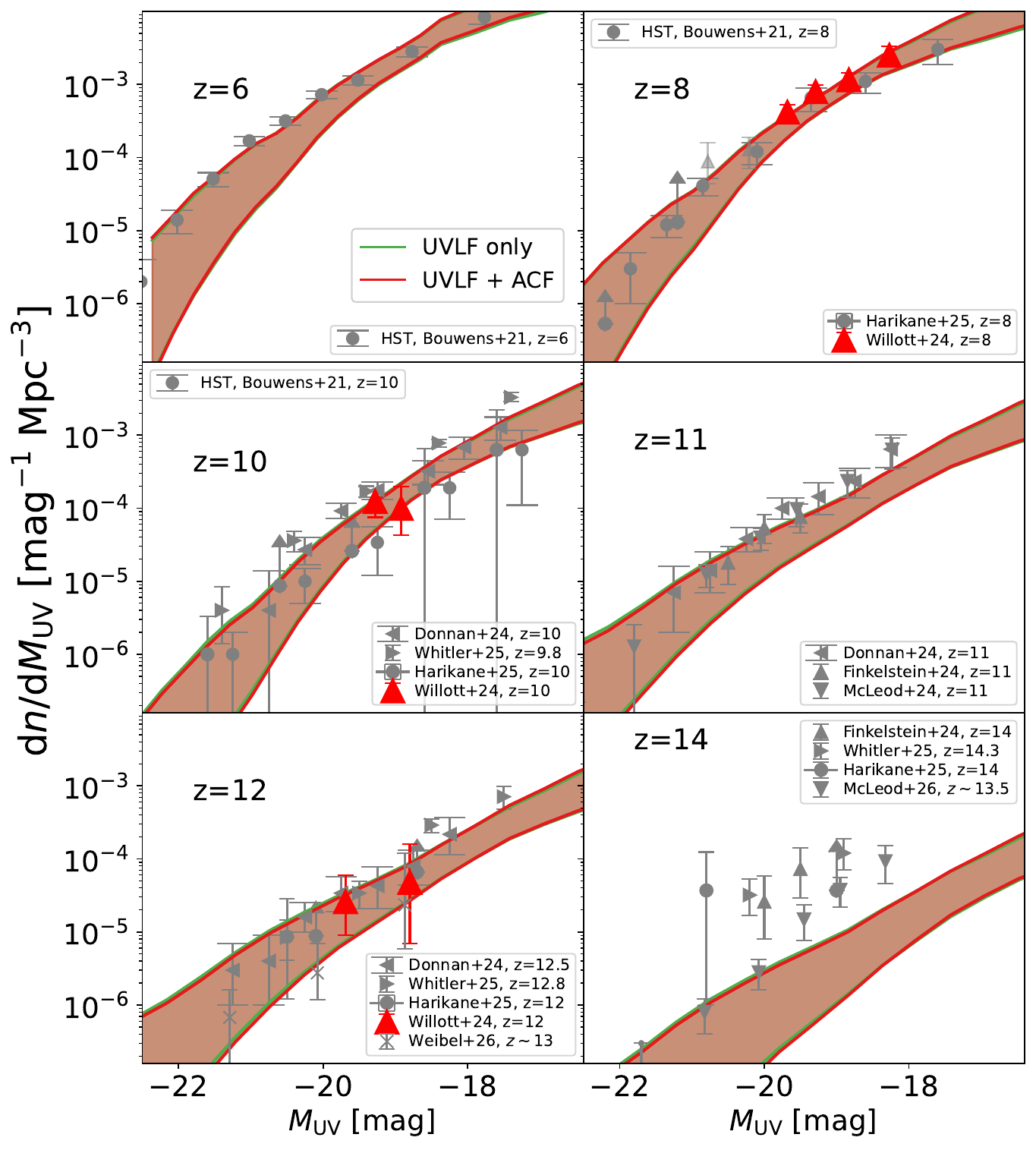}
    \caption{Posterior predictive UV LFs at $z\in\{6,8,10,11,12,14\}$, including the dust attenuation of Sect.~\ref{sec:uvlf}. The shaded regions show the $95\%$ credible intervals of the UVLF-only (green) and UVLF + ACF (red) inferences. Red triangles show the \citet{Willott2024} data used in the likelihood ($M_{\rm UV}>-20$; their brighter points are shown as light-gray triangles); other HST \citep{Bouwens2021} and JWST estimates \citep{Donnan2024, Whitler2025, Finkelstein2024, McLeod2024, Harikane2025, Weibel2025, McLeod2026} are shown in gray. Arrows mark lower and upper limits. The two posteriors nearly coincide.}
    \label{fig:uvlf_post}
\end{figure}

\subsection{The model reproduces both datasets}
\label{sec:res_uvlf}
\label{sec:res_acf}

Figure~\ref{fig:uvlf_post} shows the posterior predictive UV LFs of both inferences, with the \citet{Willott2024} data used in the likelihood as red triangles and other estimates (not used in the likelihood) in gray. The UVLF-only posterior reproduces the observed UV LFs across the fitted redshift range, $z=8$--$12$.

The UV LF posterior also generalizes beyond the red triangles used in the likelihood.  The predictions extrapolate well to both brighter and fainter magnitudes at the fitted redshifts, suggesting that the physically-motivated scaling relations hold across halo masses. The posterior at $z=6$ is consistent with the HST estimates from \citet{Bouwens2021}.  At $z=14$ the posterior is systematically below most estimates, though is reasonably consistent with the more recent conservative JWST estimates from \citet{McLeod2026}.  It is important to note that the extrapolation to higher redshift is driven both by the model and the data used in the likelihood. Indeed, if we use the UV LFs of \citet{Harikane2025} instead of \citet{Willott2024} in the likelihood (Appendix~\ref{sec:app_inf_uvlf_others}, Fig.~\ref{fig:uvlf_diff_uvlf}), the predictive posteriors at $z=14$ are consistent with \citet{McLeod2026}, but they are much broader: at $z=6$ the upper edge of the $95\%$ credible interval lies 6--50 times above the HST data, although the median stays within a factor of $\sim2$ of them.  Indeed, different observational estimates are in strong tension with each other at $z=14$. This highlights the need for a homogeneous UV LF dataset spanning a wide range of redshifts and UV magnitudes.

\begin{figure}
    \centering
    \includegraphics[width=1.0\linewidth]{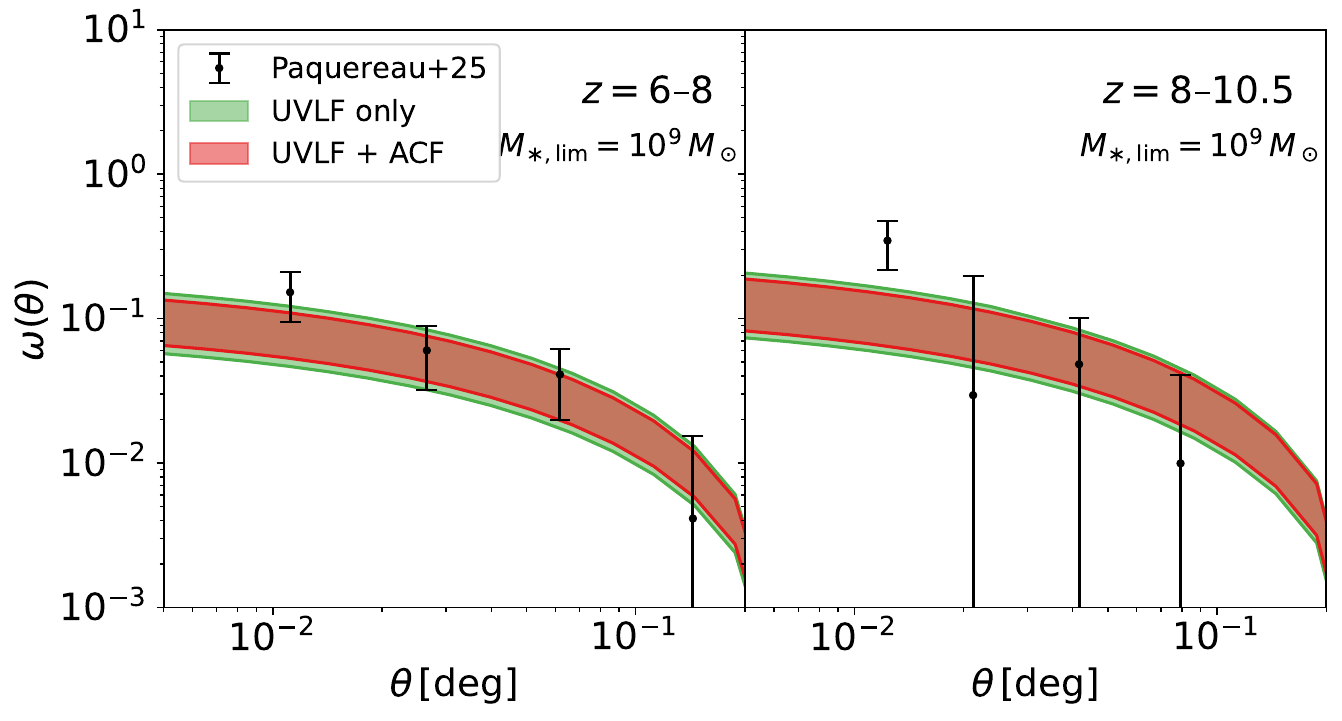}
    \caption{Posteriors of our galaxy model evaluated in the ACF space. The shaded regions show the $95\%$ credible intervals of the UVLF-only (green) and UVLF + ACF (red) inferences. The posteriors are evaluated at $z\in[6,8]$ and $z\in[8,10.5]$ bins in the left and right panel, respectively. We also show observations from \citet{Paquereau2025} as black points; points with $\omega<0$ are not shown. Adding the ACF to the likelihood narrows the $95\%$ credible interval on the fitted scales ($\theta>0.005^\circ$) by $\approx20\%$, trimming both edges.}
    \label{fig:ang_clust_post}
\end{figure}

The UVLF + ACF inference is also consistent with the observed ACFs in both redshift bins (Fig.~\ref{fig:ang_clust_post}).  Conversely, adding the clustering measurements does not significantly alter the UV LF posterior (the green and red bands in Fig.~\ref{fig:uvlf_post} closely overlap). The two datasets are therefore mutually consistent, while the current uncertainties of the clustering measurements limit their impact on the predicted UV LF.

Overall, Figs.~\ref{fig:uvlf_post} and~\ref{fig:ang_clust_post} indicate that the inferred galaxy population is broadly consistent with both JWST and pre-JWST measurements within current uncertainties.  This suggests that galaxy evolution is reasonably smooth and can be captured by extrapolations of existing scaling relations and associated scatter.  We test this claim further using model extensions in Sect.~\ref{sec:extensions}.

\begin{figure}
    \centering
    \includegraphics[width=1.0\linewidth]{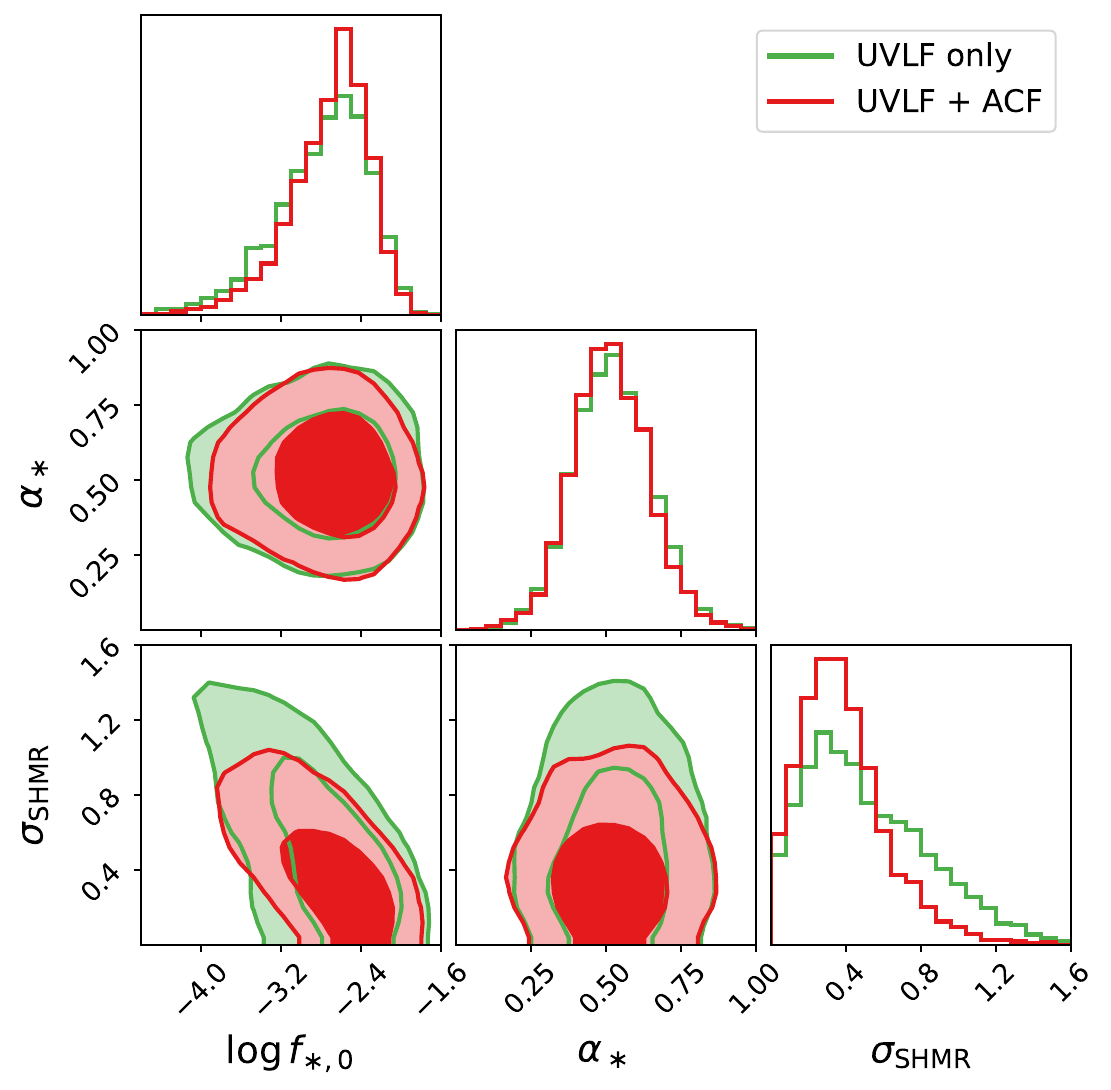}
    \caption{Posterior of the SHMR parameters from the UVLF-only (green) and UVLF + ACF (red) inferences. Contours show the 68\% and 95\% credible regions.}
    \label{fig:corner_SHMR}
\end{figure}

\subsection{UV luminosity functions and clustering are complementary}
\label{sec:res_shmr}

\begin{figure*}
    \centering
    \includegraphics[width=0.9\textwidth]{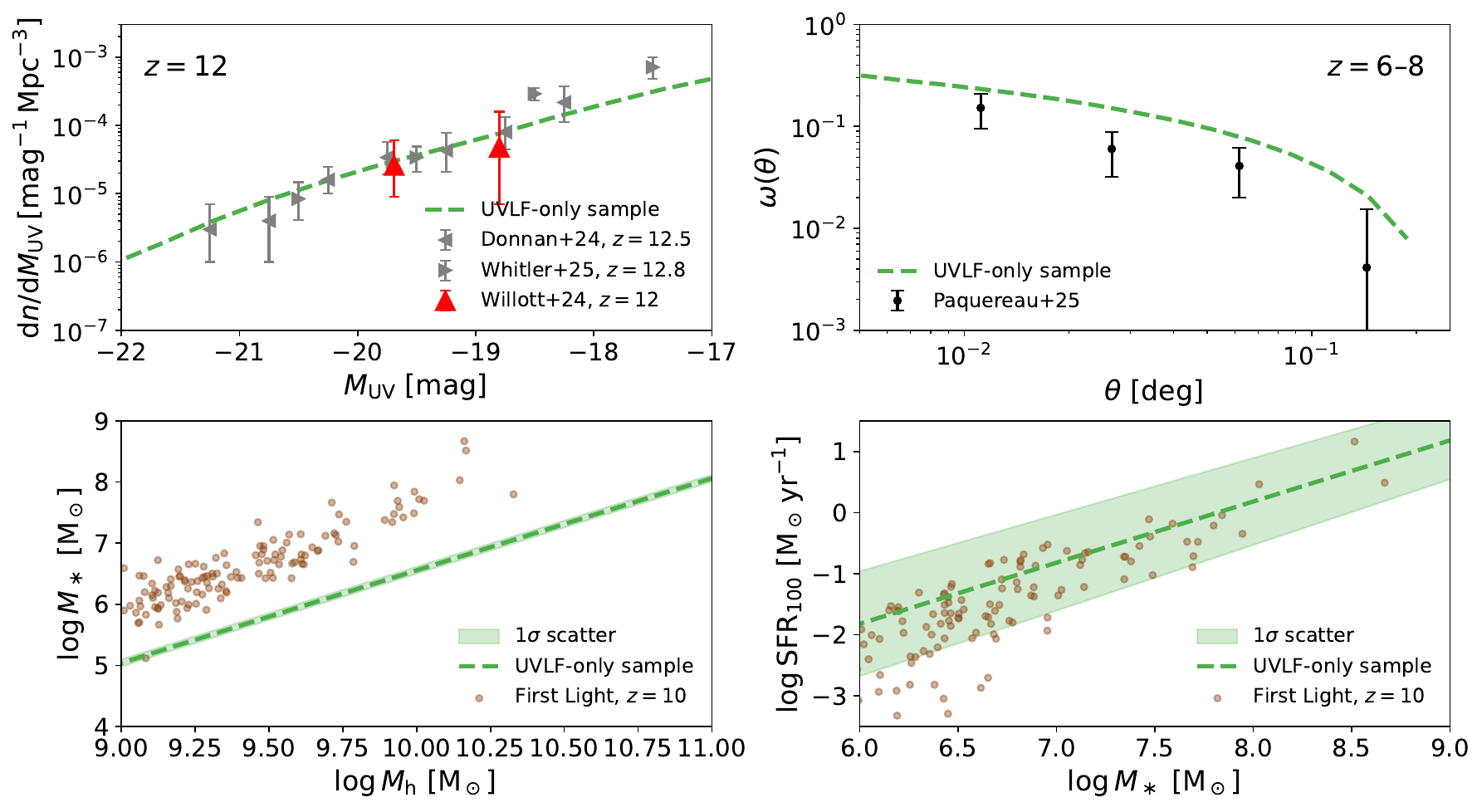}
    \caption{A model from the UVLF-only posterior that fits the UV LF but not the clustering. \textit{Top left:} UV LF at $z=12$ of the model (green dashed) and observational estimates \citep[gray;][]{Donnan2024, Whitler2025}, with the \citet{Willott2024} data used in the likelihood in red. \textit{Top right:} ACF, $\omega(\theta)$, of galaxies with $M_\ast>10^9\,M_\odot$ at $z=6$--$8$, compared with \citet{Paquereau2025} (black). \textit{Bottom left:} SHMR of the model, mean (dashed) and $1\sigma$ scatter (shaded), compared, for visual reference, with First Light galaxies at $z=10$ \citep[brown;][]{Ceverino2018}. \textit{Bottom right:} the same for the SFMS. The low SHMR is compensated by a broad SFMS, so the model reproduces the UV LF while overpredicting the mass-complete ACF (which is only sensitive to the SHMR).}
    \label{fig:ang_elp}
\end{figure*}

The UV LF and the ACF constrain different combinations of the SHMR parameters, as was shown in Sect.~\ref{sec:sensitivity}.  This complementarity was  seen in the posterior predictive ACFs (Fig.~\ref{fig:ang_clust_post}). The UVLF-only posterior predicts a wider range of clustering amplitudes, reflecting the degeneracy between galaxy–halo connection parameters when only abundance information is used.
Including the clustering measurements from \citet{Paquereau2025} narrows the $95\%$ credible interval of the amplitude on the fitted scales by $\approx20\%$, excluding both high-amplitude models (as in Fig.~\ref{fig:ang_elp}) and low-amplitude ones with a large SHMR scatter.

To further quantify this trend, in Fig.~\ref{fig:corner_SHMR} we show the posteriors of the three free SHMR parameters from both the UVLF-only and UVLF + ACF inferences.  Adding the ACF to the likelihood shrinks the $68\%$ credible interval of $\sigma_{\rm SHMR}$ from $[0.20,0.86]$ to $[0.16,0.60]$, while that of $\log f_{\ast,0}$ barely changes.  Indeed, referring to Fig.~\ref{fig:sensitivity} we can see that $\sigma_{\rm SHMR}$, which controls the scatter around the mean SHMR, is the only parameter for which the constraining power of the mass-selected ACFs is comparable to the UV LFs. This constraint comes from the $10^9\,M_\odot$ sample, the lowest stellar-mass limit for which our model fits the clustering well (Sect.~\ref{sec:acf}); the poorly fit $10^{8.7}\,M_\odot$ and the noisier $10^{9.3}\,M_\odot$ samples do not reduce the scatter.

Figure~\ref{fig:ang_elp} illustrates this complementarity with one model drawn from the UVLF-only posterior. It matches the target UV LF (red), but overpredicts the ACF by a factor of $\approx1.5$--$2.5$ at $\theta\lesssim0.06^\circ$. Its SHMR lies $\approx0.9$\,dex below the First Light galaxies (lower left panel), but the UV LF is recovered by compensating the low SHMR with high scatter around the SFMS ($\sigma_{\rm SFMS,0}=0.57$; lower right panel). 
In this particular model, galaxies above $M_{\ast,\rm lim}=10^9\,M_\odot$ reside only in rare, massive halos, which cluster too strongly and are disfavored by the observed ACFs. 
\begin{figure}
    \centering
    \includegraphics[width=1.0\linewidth]{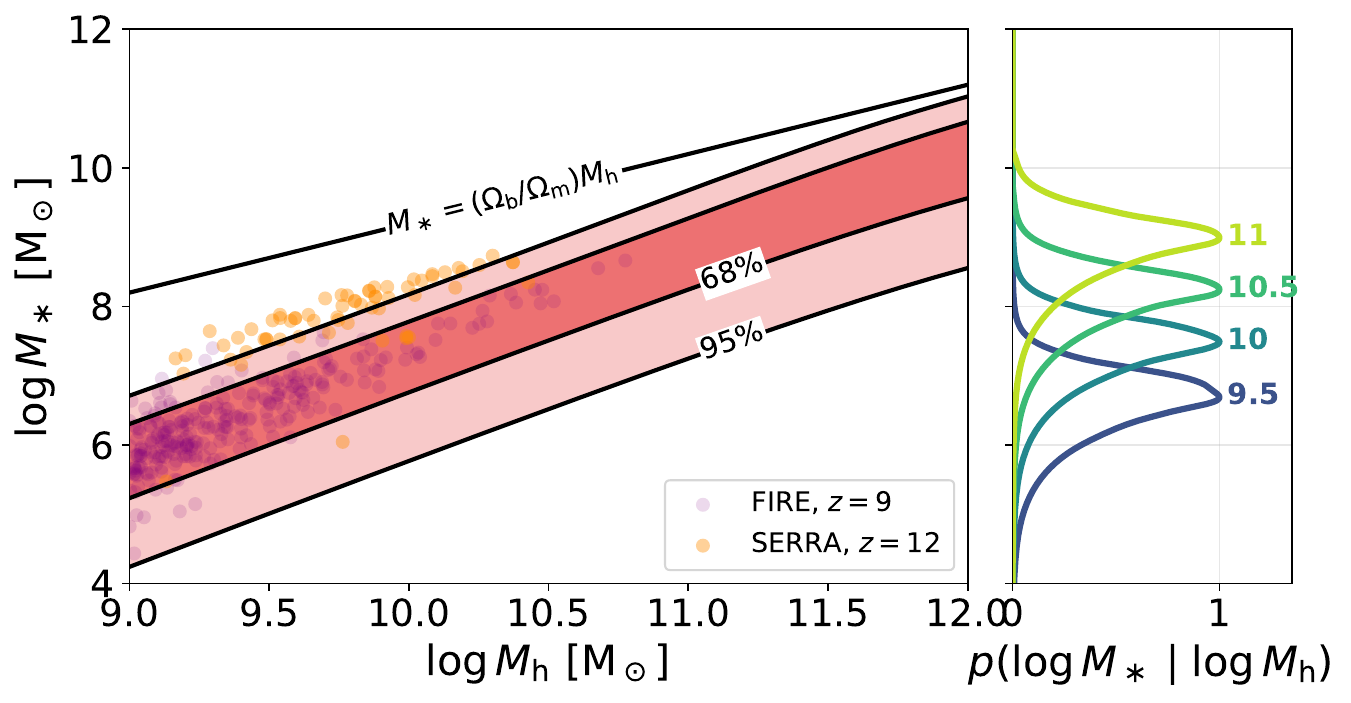}
    \caption{SHMR implied by the UVLF + ACF posterior. \textit{Left panel:} stellar mass as a function of halo mass; the shaded regions show the 68\% and 95\% credible intervals, including the uncertainties on both the mean and scatter around the mean. The black line denotes the baryon-fraction limit $M_\ast=f_{\rm b}M_{\rm h}$. Points show individual galaxies from FIRE at $z=9$ \citep[purple;][]{Ma2017} and SERRA at $z=12$ \citep[orange;][]{Pallottini2022}. \textit{Right panel:} conditional distributions of stellar mass at fixed halo mass, normalized to unit peak and labeled by their $\log(M_{\rm h}/M_\odot)$.}
    \label{fig:shmr_post}
\end{figure}

Figure~\ref{fig:shmr_post} shows the SHMR in physical space. We show only the result for the joint posterior with shaded regions, reflecting both parameter uncertainty and intrinsic scatter. In the right panel, we also show the conditional distributions for fixed $M_{\rm h}$. The FIRE galaxies lie within the $68\%$ band of the inferred SHMR and the SERRA galaxies above it. The MAP estimate of the SHMR is

\begin{equation}
    M_{\ast} = \frac{0.035}{\left(\frac{M_{\rm h}}{2\times10^{12}\,M_{\odot}}\right) ^ {-0.51} + \left(\frac{M_{\rm h}}{2\times10^{12}\,M_{\odot}}\right) ^ {0.61}} M_{\rm h},
\end{equation}
with $\sigma_{\rm SHMR} = 0.28$; the relation corresponds to $\log f_{\ast,0}=-2.63$ and $\alpha_\ast=0.51$ in Eq.~\eqref{eq:SHMR}.

\begin{figure}
    \centering
    \includegraphics[width=1.0\linewidth]{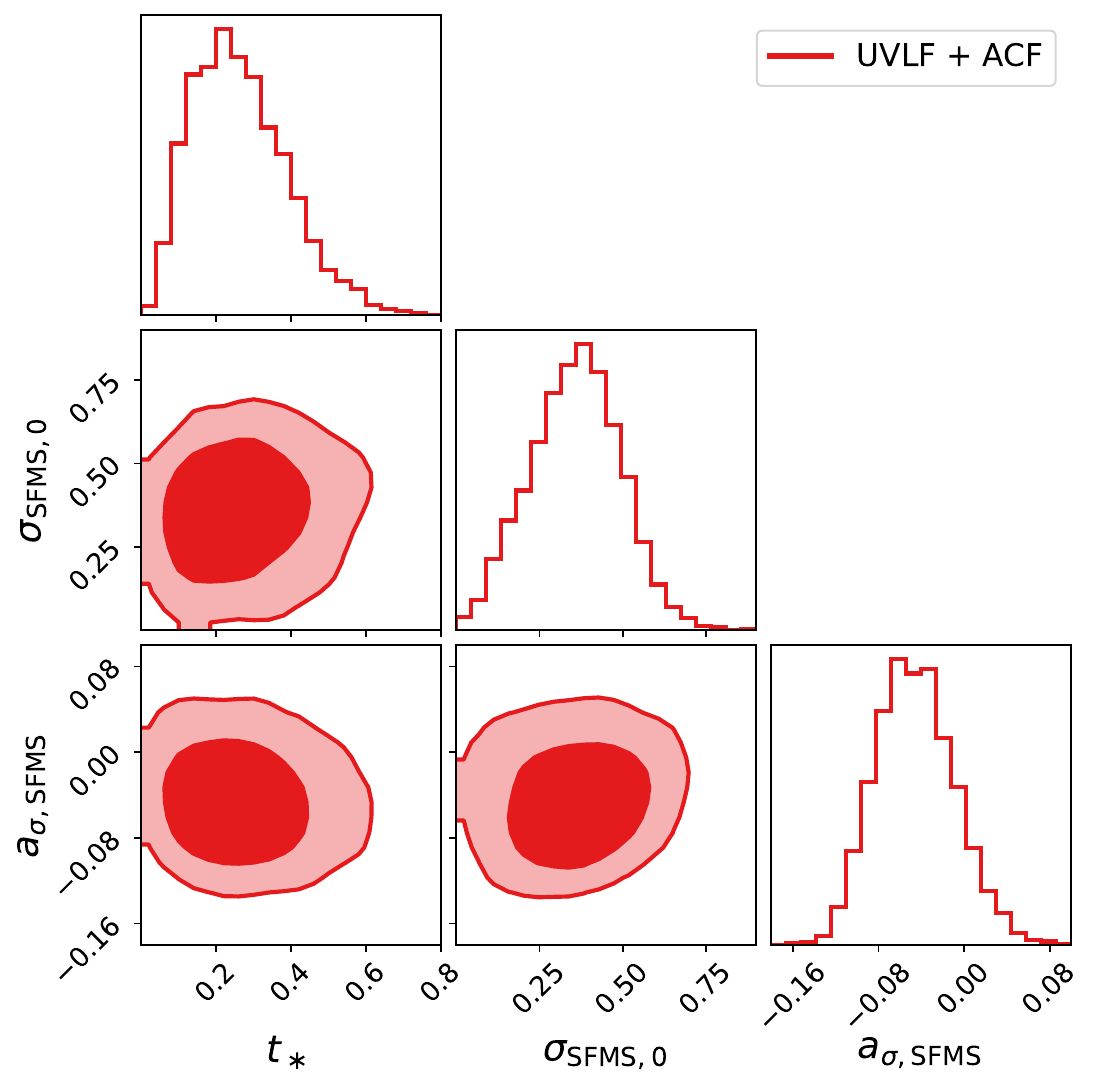}
    \caption{Posterior of the SFMS parameters from the UVLF + ACF inference; contours show the 68\% and 95\% credible regions. The bottom right panel shows the slope of the SFMS scatter, $a_{\sigma,\rm SFMS}$, which is negative at 89\% probability.}
    \label{fig:sfms_corner}
\end{figure}

\subsection{The star-forming main sequence}
\label{sec:res_sfms}

In Fig.~\ref{fig:sfms_corner} we show the posterior of the parameters determining the SFMS.  We see that the scatter around the SFMS increases towards lower stellar masses: its slope is negative, $a_{\sigma,\rm SFMS}<0$, at $89\%$ posterior probability (bottom right panel of Fig.~\ref{fig:sfms_corner}). This trend was already in our simulation-ensemble-based priors, with $83\%$ of the prior mass having negative values of $a_{\sigma,\rm SFMS}$. The trend is stronger with the other UV LF datasets of Appendix~\ref{sec:app_inf_uvlf_others}: $P(a_{\sigma,\rm SFMS}<0)=0.91$, $0.99$ and $0.997$ for \citet{Harikane2025}, \citet{Whitler2025} and \citet{Donnan2024}, respectively.  All three SFMS parameters are constrained by the UV LF.  The ACF does not depend on them (Sect.~\ref{sec:sensitivity}) and affects them only through their degeneracies with the SHMR in the UV LF.

The MAP value corresponds to a star-formation timescale of $\approx0.9\,t_{\rm dyn}(z)$, or $\approx140$--$190$\,Myr at $z=8$--$10$. This is lower than some previous estimates \citep{Park2019, Nikolic2023} and indicates that stellar assembly in galaxies at $z\gtrsim8$ proceeds rapidly, as predicted by the simulations.

\begin{figure}
    \centering
    \includegraphics[width=1.0\linewidth]{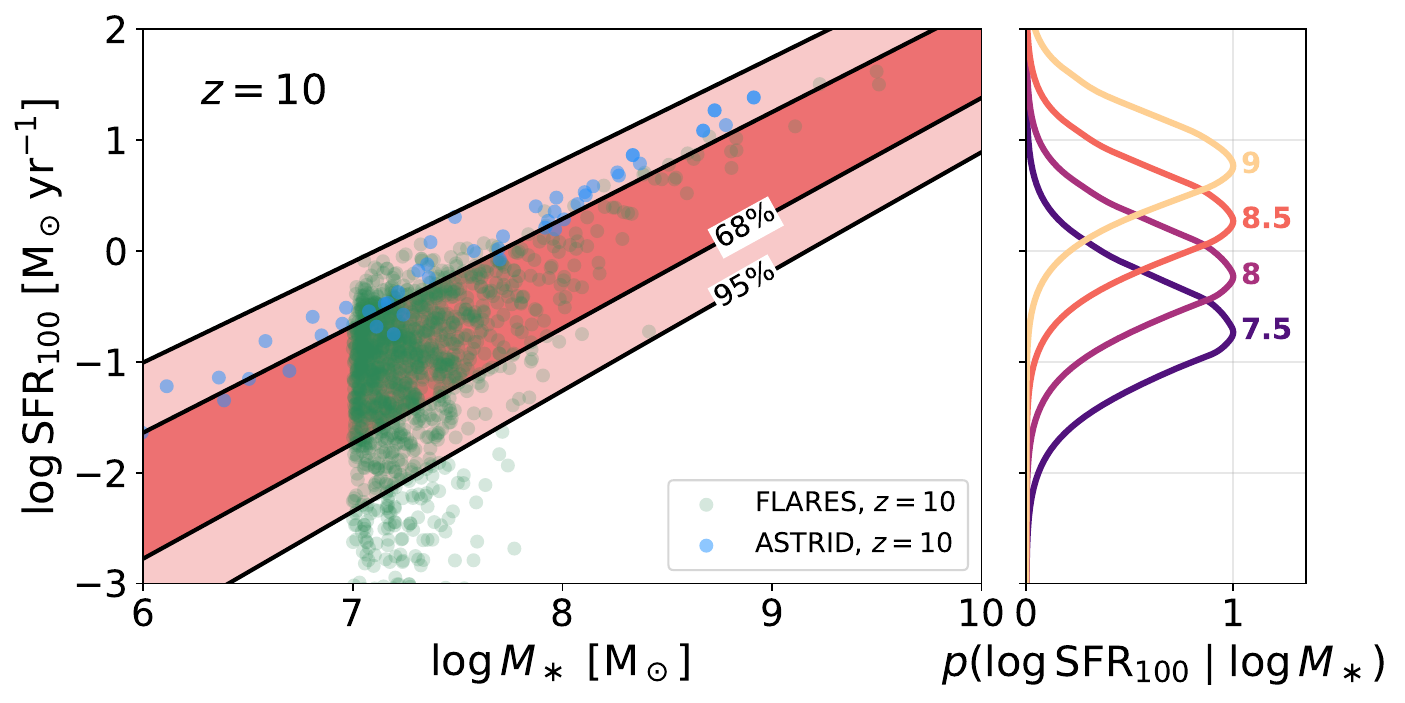}
    \caption{SFMS implied by the UVLF + ACF posterior at $z=10$. \textit{Left panel:} $\log{\rm SFR}_{100}$ as a function of stellar mass; the shaded regions show the 68\% and 95\% credible intervals, including both parameter uncertainty and intrinsic scatter. Points show individual galaxies from FLARES \citep[green;][]{Lovell2021} and ASTRID \citep[blue;][]{Bird2022} at $z=10$. \textit{Right panel:} conditional distributions of SFR$_{100}$ at fixed stellar mass, normalized to unit peak and labeled by $\log(M_\ast/M_\odot)$.}
    \label{fig:sfms_post}
\end{figure}

In Fig.~\ref{fig:sfms_post} we show the posterior of the SFMS in the physical space with the total uncertainty encompassing the intrinsic scatter and the uncertainty on the distribution.  The galaxy population inferred from the JWST data can be placed directly among the predictions from the simulations used to construct the prior. Each MAP parameter lies within the range spanned by the five simulations (Fig.~\ref{fig:prior_corner}), although no single simulation matches all of them.  The ASTRID galaxies lie above the median of the inferred SFMS and the FLARES galaxies scatter around and below it (Fig.~\ref{fig:sfms_post}).

For reference, the SFMS of the MAP model is
\begin{equation}
    {\rm SFR}_{100} = \frac{M_{\ast}}{0.20\,H^{-1}(z)} = \frac{M_{\ast}}{0.92\,t_{\rm dyn}(z)},
    \label{eq:sfms_map}
\end{equation}
\noindent and its mass-dependent scatter is
\begin{equation}
    \sigma_{\rm SFMS} = \begin{dcases}
    -0.064\,\log\left(\frac{M_{\ast}}{10^{10}\,M_{\odot}}\right) + 0.35,& \text{if } M_{\ast}\leq 10^{10}\,M_{\odot}\\
    0.35,              & \text{otherwise}
    \label{eq:sigma_sfms}
\end{dcases}.
\end{equation}

\begin{figure}
    \centering
    \includegraphics[width=1.0\linewidth]{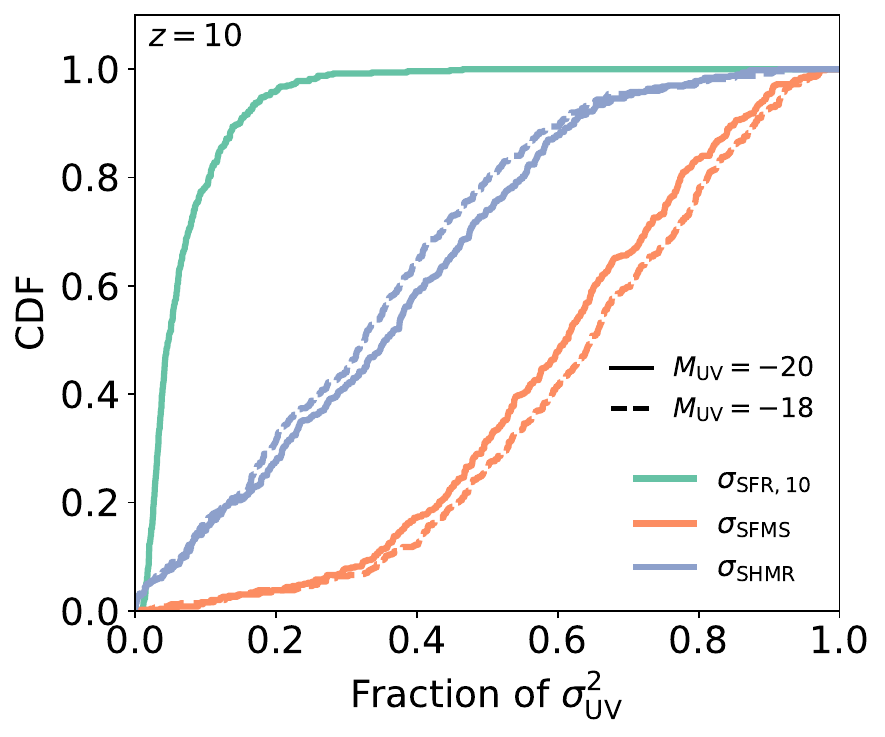}
    \caption{Contribution of different terms to the total variance in UV magnitude at a fixed halo mass at $z=10$. Colored curves show the cumulative distributions, over the UVLF + ACF posterior, of the fractional contribution of each source of scatter to the total $\sigma_{\rm UV}^2$: excess burstiness (green), SFMS scatter (orange) and SHMR scatter (blue). Solid lines are evaluated at $M_{\rm UV}=-20$; dashed lines show the SFMS and SHMR contributions at $M_{\rm UV}=-18$ (the SFR$_{10}$ contribution is omitted at $M_{\rm UV}=-18$ to avoid crowding, and is small at both magnitudes). The SFR$_{10}$ curve rises almost immediately because its contribution is below $15\%$ in $90\%$ of posterior samples. 
    In the median, the SFMS scatter contributes most of the variance, with a notable contribution from the SHMR scatter.   
The SFMS contribution increases for fainter magnitudes (solid vs dashed lines), consistent with the mass-dependence of $\sigma_{\rm SFMS}$.}
    \label{fig:frac_sigma}
\end{figure}

\subsection{Scatter in the SFMS is most important in setting the stochasticity of UV magnitudes}
\label{sec:res_stochasticity}

The large SFMS scatter, rising towards low masses, is also the main source of stochasticity in the UV magnitudes of galaxies at fixed halo mass, which has been invoked to explain the abundance of bright galaxies at $z\gtrsim10$ (Sect.~\ref{sec:galfor}). Our inference framework allows us to decompose this total stochasticity into its individual physical contributions \citepalias{Nikolic24}. We obtain the total scatter, $\sigma_{\rm UV}$, by fitting a Gaussian to the normalized $p(M_{\rm UV}\mid M_{\rm h})$ at $z=10$, at the halo mass at which its mean is $M_{\rm UV}=-20$ ($-18$), i.e., $M_{\rm h}\approx10^{10.9}$ ($10^{10.4}$)\,$M_\odot$ for the posterior median. To isolate the contribution of each source of scatter, we recompute $\sigma_{\rm UV}$ after suppressing one source at a time, setting its scatter ($\sigma_{\rm SHMR}$; $\sigma_{\rm SFMS,0}$ and $a_{\sigma,\rm SFMS}$; or $\sigma_{\rm SFR,10}$) to a negligible value while keeping all other parameters at their posterior sample values. The share of the variance due to each source is then $1-(\sigma_{\rm UV,\,reduced}/\sigma_{\rm UV,\,full})^2$. As the sources add in quadrature, these shares add up to unity to within a few percent. Figure~\ref{fig:frac_sigma} shows their distribution over the UVLF + ACF posterior.

At $M_{\rm UV}=-20$, the SFMS scatter accounts for a median $61\%$ of the variance, against $36\%$ for the SHMR scatter, and it is the larger of the two in $72\%$ of the posterior samples.  Therefore the SFMS scatter is generally more important, contributing, in the median, about $1.7$ times as much as the SHMR scatter to the total UV luminosity stochasticity at fixed halo masses. Because $\sigma_{\rm SFMS}$ increases towards lower stellar masses, its share rises to $65\%$ at $M_{\rm UV}=-18$. Additional burstiness on short ($\sim10$\,Myr) timescales contributes little ($\approx5\%$), as UV luminosity is primarily sensitive to star formation averaged over longer timescales ($\sim100$\,Myr). Since the data do not constrain $\sigma_{\rm SFR,10}$, this contribution follows from its prior, which lies at the lower end of observational estimates (Sect.~\ref{sec:priors}); for $\sigma_{\rm SFR,10}\approx0.4$--$0.5$\,dex it would rise to $\approx15$--$25\%$, still below the SFMS contribution. Within this range, models invoking enhanced short-timescale burstiness alone are therefore unlikely to significantly affect the UV luminosity function. 

Despite the relative claim that SFMS scatter is typically more important than the SHMR scatter, the individual fractions are only weakly constrained, with $68\%$ credible intervals of $[39\%,81\%]$ for $\sigma_{\rm SFMS}$ and $[11\%,57\%]$ for $\sigma_{\rm SHMR}$, whereas their total is better determined ($\sigma_{\rm UV}=1.36^{+0.36}_{-0.39}$\,mag at $M_{\rm UV}=-20$, $68\%$ credible interval). Additional probes are necessary to further clarify the role of each source of stochasticity.

\begin{figure}
    \centering
    \includegraphics[width=1.0\linewidth]{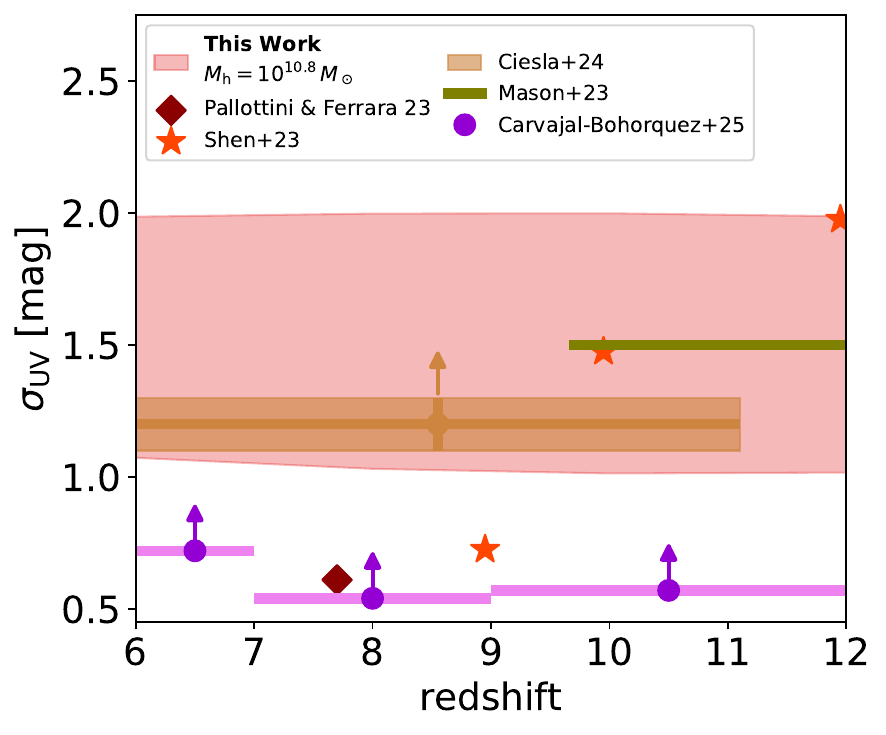}
    \caption{Redshift evolution of the total scatter in UV magnitude, $\sigma_{\rm UV}$ (in mag), at a halo mass of $10^{10.8}\,M_\odot$ (red band: 68\% credible interval of our posterior), compared with other theoretical works \citep{Pallottini2023, Mason2023, Shen2023} and with observational estimates at fixed stellar mass, which are lower limits on $\sigma_{\rm UV}$ at fixed halo mass \citep{Ciesla2024, Carvajal-Bohorquez2025}.}
    \label{fig:sigma_z}
\end{figure}

Having quantified the origin of the scatter, we next examine its evolution with redshift (Fig.~\ref{fig:sigma_z}). This enables direct comparison with previous studies that invoke large values of $\sigma_{\rm UV}$ to explain the overabundance of bright galaxies in JWST photometric samples \citep{Mason2023, Shen2023}, as well as with works that estimate this quantity from simulations \citep{Pallottini2023} or observations \citep{Ciesla2024, Carvajal-Bohorquez2025}. Our $\sigma_{\rm UV}$ hardly evolves between $z=6$ and 12, reflecting the redshift-independent scatters of our fiducial model (Sect.~\ref{sec:extensions}), and is consistent with the values required to reproduce the observed UV luminosity functions. Estimates from SED fitting of individual galaxies give a similar \citep{Ciesla2024} or smaller \citep{Carvajal-Bohorquez2025} scatter, but the comparison is not like-to-like: such measurements probe $p(M_{\rm UV} \mid M_\ast)$ rather than $p(M_{\rm UV} \mid M_{\rm h})$, and so offer insight into only part of the stochasticity budget rather than the full one modeled here. Moreover, the unknown completeness of these samples could lead them to underestimate the total scatter. 

\subsection{The data do not require extensions of the fiducial model}
\label{sec:extensions}
\begin{figure*}
    \centering
    \includegraphics[width=1.0\linewidth]{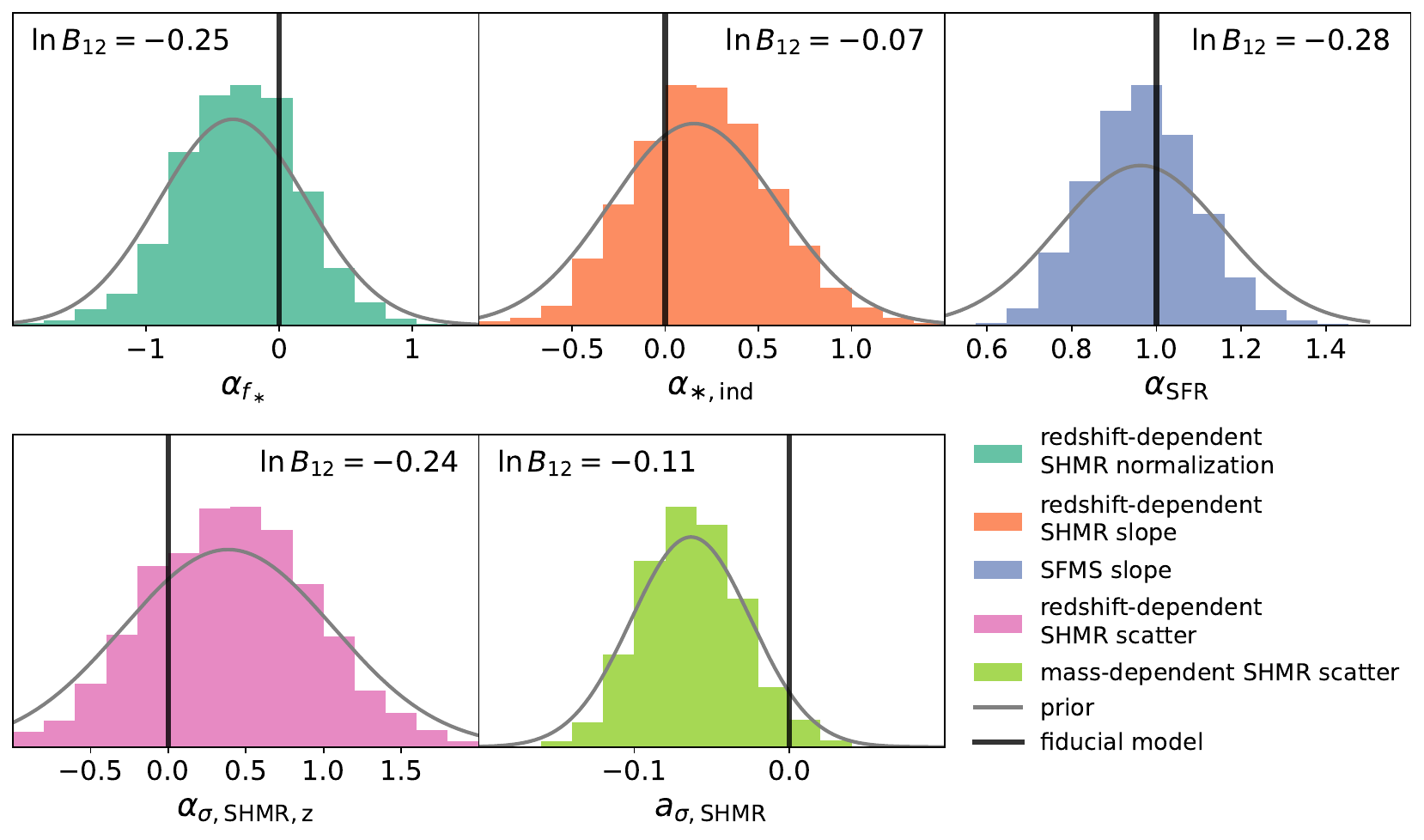}
    \caption{Posterior of the added parameter of each model extension (histograms), with its simulation prior (gray) and fiducial value (black line). Top row, left to right: redshift-dependent SHMR normalization, redshift-dependent SHMR slope, SFMS slope; bottom row: redshift- and mass-dependent SHMR scatter. $\ln B_{12}$ is the log Bayes factor of the extension relative to the fiducial model.}
    \label{fig:alternatives}
\end{figure*}

Our fiducial model rests on simplifying assumptions that could break down at high redshift: an SHMR whose mean and scatter are independent of redshift, with a scatter that is also independent of halo mass (Sect.~\ref{sec:shmr}), and an SFR that scales linearly with stellar mass along the SFMS (Eq.~\eqref{eq:sfms}). We test these assumptions with five physically motivated extensions, each adding one free parameter, with a prior obtained from the simulations as in Sect.~\ref{sec:priors}:

\begin{itemize}
    \item \textbf{Redshift dependent SHMR normalization:} While the SHMR is traditionally believed to be redshift independent, potential redshift evolution is found in some inference works \citep{Shuntov2022, Paquereau2025, Shuntov2025}. The evolution can be caused by rapid evolution in the strength of feedback, inflows/outflows, mergers, or metallicity effects \citep[see, e.g.,][]{Wechsler2018}. We parametrize it through $f_\ast (z) = f_{\ast,0}\,\left( \frac{1+z}{11}\right)^{\alpha_{f_\ast}}$.
    \item \textbf{Redshift dependent SHMR slope:} Similarly, the slope of the SHMR can evolve due to the efficiency of feedback at high redshifts \citep[e.g.,][]{Shuntov2025}. We parametrize potential redshift evolution as: $\alpha_\ast (z) = \alpha_{\ast,0} + \alpha_{\ast, \rm ind}\,\left( \frac{1+z}{11} - 1\right)$, so that $f_{\ast,0}$ and $\alpha_{\ast,0}$ are the values at $z=10$.
    \item \textbf{Change to the slope of the SFMS:} The slope of the SFMS can depart from unity in the limit in which these same feedback mechanisms affect accretion of new material \citep[e.g.,][]{diCesare2026, Merida2026}. We parametrize the SFMS using a more general form in this case as: ${\rm SFR}_{100} = \frac{M_{\ast}}{t_{\ast} H^{-1}(z)} \left( \frac{M_{\ast}}{10^9\,M_\odot}\right)^{\alpha_{\rm SFR}-1}$
    \item \textbf{Redshift dependent SHMR scatter:} Our fiducial model assumes the SHMR scatter to be redshift-independent (Sect.~\ref{sec:shmr}); we test this assumption by allowing it to evolve linearly with redshift: $\sigma_{\rm SHMR} (z) = \sigma_{\rm SHMR,0} \left[1 + \alpha_{\sigma,\rm SHMR,z} \left( \frac{1+z}{12} - 1\right)\right]$, so that $\sigma_{\rm SHMR,0}$ is the scatter at $z=11$.
    \item \textbf{Mass dependent SHMR scatter:} Similarly, we test whether the SHMR scatter depends on halo mass, using the same piecewise functional form already adopted for the SFMS scatter (Sect.~\ref{sec:sfr}) and anchored at the characteristic mass, $M_{\rm knee}$, at which the mean SHMR bends: $\sigma_{\rm SHMR} = \sigma_{\rm SHMR,0} + a_{\sigma,\rm SHMR}\,\min\left[\log\left(M_{\rm h}/M_{\rm knee}\right),\,0\right]$.
\end{itemize}

For each extension, we repeat the UVLF + ACF inference. Figure~\ref{fig:alternatives} shows the posterior of each added parameter, together with its simulation-based prior (gray curves), its fiducial value, which corresponds to no mass or redshift dependence ($\alpha_{\rm SFR}=1$ for the SFMS slope, zero otherwise), and the Bayes factor of each extension relative to the fiducial model.

To assess whether the data favor these extensions, we compare each posterior with its prior and compute the Bayes factor, i.e., the ratio of the Bayesian evidence of the extended model to that of the fiducial model. The posteriors closely follow the simulation priors: their medians lie within $0.1$ prior standard deviations of the prior mean and they are only $\approx15$--$35\%$ narrower, and the fiducial value lies within the $95\%$ credible interval in all five cases (at its edge for the mass-dependent scatter, whose simulation prior already favors a scatter increasing towards lower masses). All Bayes factors are small, $-0.28\leq\ln B_{12}\leq-0.07$ (uncertainties $\approx0.08$), so the current data cannot distinguish the extensions from the fiducial model. Additional probes and future data will allow these extensions to be tested more rigorously.

Interestingly, using a different galaxy model and information criteria (Akaike, Bayesian and deviance; applied to UV LFs spanning $z=4$--16, without clustering information), \citet{Kar2026} find that redshift evolution \textit{is} required in the low-mass slope of their star-formation-efficiency--halo-mass relation (directly analogous to our $\alpha_\ast$), while their normalization and UV-magnitude scatter remain redshift-independent, consistent with our own findings for those parameters. This tension in the slope result may reflect the narrower redshift range of our UV LF data ($z\approx8$--12, vs.\ 4--16), differences in model-selection philosophy---Bayesian evidence informed by simulation-motivated priors versus a fixed per-parameter penalty---the absence of clustering constraints in their analysis, or degeneracies with other redshift-evolving quantities that are broken differently once clustering is included. Within the redshift range of our data, however, the fiducial model suffices.

\subsection{The model disfavors most photometric candidates at $z>14$}
\label{sec:res_highz}

\begin{figure}
    \centering
    \includegraphics[width=1.0\linewidth]{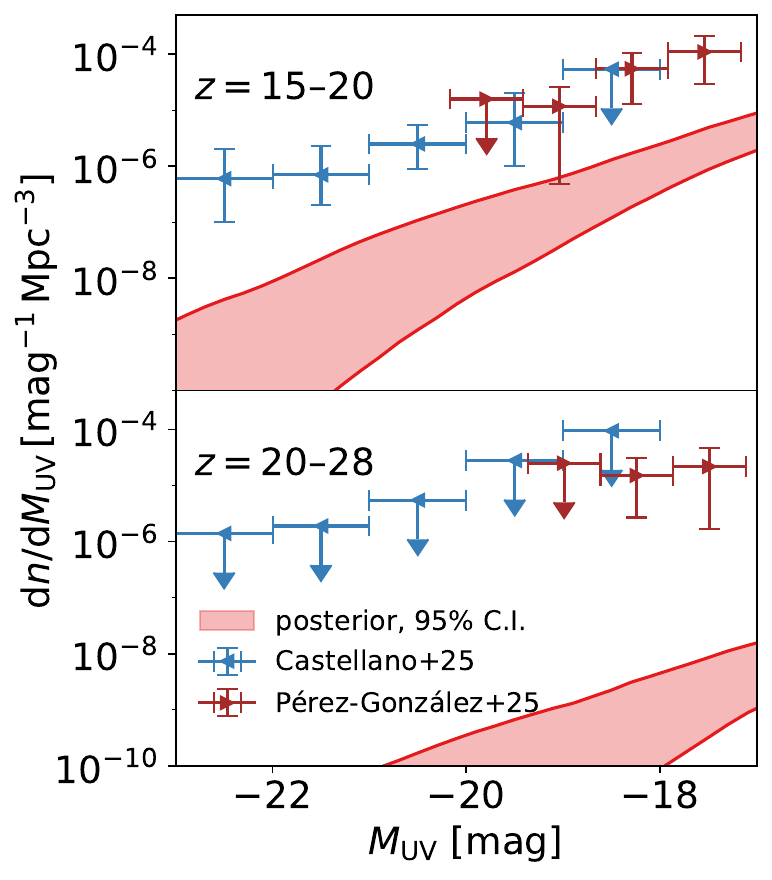}
    \caption{Posterior predictive UV luminosity function averaged over $z=15$--$20$ (top) and $z=20$--$28$ (bottom; red bands: $95\%$ credible interval), compared to estimates from the photometric candidate samples of \citet{Castellano2025} ($z=15$--$20$ and $20$--$28$) and \citet{Perez_Gonzalez2025} ($z=16$--$20$ and $24$--$26$). Our fiducial posterior lies orders of magnitude below the observational estimates from candidates. The bins of \citet{Perez_Gonzalez2025} typically hold two candidates, so the gap remains even if only one per bin were genuine.}
    \label{fig:superhigh}
\end{figure}

Finally, we use the inferred model to make predictions at higher redshifts, $z>14$, where current observations are limited to photometric candidates. Recent studies \citep{Perez_Gonzalez2025, Castellano2025} have identified photometric candidates consistent with a Lyman-break at $z>14$. While none of these sources are yet spectroscopically confirmed, these works have presented tentative estimates of the UV luminosity function based on the candidate samples. 

In Fig.~\ref{fig:superhigh}, we compare the posterior predictive UV luminosity function to the estimates derived from these photometric samples. The predicted UV LF lies well below the reported estimates, by orders of magnitude across the luminosity range, even if only one candidate per bin were genuine. Inferences using the UV LFs of \citet{Donnan2024} or \citet{Whitler2025} instead give similar predictions; only the much broader posterior from the \citet{Harikane2025} UV LFs reaches the \citet{Castellano2025} estimates at $z=15$--$20$, although its median remains more than an order of magnitude below them.   This suggests that a scenario in which a significant fraction of the photometric candidates lie at $z>14$ is disfavored within our framework. If these candidates were confirmed, they would necessitate significant extensions of our fiducial model, such as a strong redshift-dependence of the SHMR and SFMS (see, e.g., \citealt{Yung2025}), more extreme SFHs and/or IMF evolution, or even more exotic sources, such as accreting primordial black holes (e.g., \citealt{Matteri2025}) or pair-instability supernovae (e.g., \citealt{Ferrara2026, Jeon2026}).

%%%%%%%%%%%%%%%%%%%%%%%%%%%%%%%%%%%%%%%%%%%%%%%%%%%%%%%%%%%%%%

\section{Conclusions}
\label{sec:conclusion}
JWST has found more UV-bright galaxies at $z\gtrsim10$ than were expected before its launch. The proposed explanations, from a higher efficiency to an enhanced stochasticity of star formation, cannot be distinguished using UV LFs alone, and comparing them is further complicated by the fact that different galaxy models describe galaxy formation with different, model-specific parameters.

To address this, we extended the model of \citetalias{Nikolic24} into a ``universal language'' for galaxy populations: a chain of conditional distributions linking halo mass, stellar mass, star formation rates averaged over $100$ and $10$\,Myr, metallicity and UV magnitude. Fitting this model to five hydrodynamic simulations, we constructed a broad prior that spans their predictions. We then inferred the model parameters from the UV LFs of \citet{Willott2024} at $z\approx8$--12 and the ACFs of mass-selected galaxies of \citet{Paquereau2025} at $z\approx6$--10.5, forward-modeling both observables.

Once expressed in this language, the five simulations describe very different galaxy populations: their posteriors are narrow, but disjoint in the joint parameter space. The SHMR and SFMS inferred from JWST data lie within the range spanned by the simulations, although no single simulation matches them.

UV LFs and clustering provide complementary information. The ACF responds to a different combination of the SHMR normalization and scatter than the UV LF, and is insensitive to the SFMS. Adding the mass-complete clustering sample disfavors a large scatter around the SHMR, $\sigma_{\rm SHMR}=0.35^{+0.25}_{-0.19}$\,dex, compared with $0.45^{+0.41}_{-0.26}$\,dex from the UV LFs alone, while leaving the UV LF posterior nearly unchanged.

With simulation-based priors, the scatter around the SFMS increases towards lower stellar masses ($89\%$ posterior probability, compared to $83\%$ under the prior), as predicted by the simulations. The mean SFMS corresponds to a star-formation timescale of about one halo dynamical time, implying rapid stellar assembly at $z\gtrsim8$.

The SFMS scatter is also the main source of stochasticity in UV magnitude at fixed halo mass, accounting for a median $\sim61\%$ of the total variance at $M_{\rm UV}=-20$, against $\sim$36\% for the SHMR scatter.  Bayesian model comparison shows that the current data do not require a redshift-dependent SHMR, a redshift- or mass-dependent SHMR scatter, or an SFMS slope different from unity ($|\ln B_{12}|<0.3$). Extrapolated beyond the fitted range, our model under-predicts most UV LF estimates at $z\approx14$ and lies orders of magnitude below the UV LFs derived from photometric candidates at $z>14$, disfavoring a scenario in which a significant fraction of these candidates lie at such redshifts. Reconciling the two would require revised candidate samples or substantial changes to the model.

More generally, our universal language can serve as an interface between theoretical models and observations. In future work, we will extend the framework to additional observables, such as nebular-line emission that probes star formation on shorter timescales, and add dust attenuation as a further conditional distribution to model the bright end of the UV LF.

%%%%%%%%%%%%%%%%%%%%%%%%%%%%%%%%%%%%%%%%%%%%%%%%%%%%%%%%%%%%%%
\begin{acknowledgements}
IN acknowledges support by European Union ERC grant RISES (101163035). The Cosmic Dawn Center (DAWN) is funded by the Danish National Research Foundation under grant DNRF140. Views and opinions expressed are those of the author(s) only and do not necessarily reflect those of the European Union or the European Research Council. Neither the European Union nor the granting authority can be held responsible for them. The Tycho supercomputer hosted at the SCIENCE HPC center at the University of Copenhagen was used for supporting this work. We gratefully acknowledge computational resources of the Center for High Performance Computing (CHPC) at SNS. AM acknowledges support from the Italian Ministry of Universities and Research (MUR) through the PRIN project ``Optimal inference from radio images of the epoch of reionization''. 
SGM has received funding from the European Union’s Horizon 2020 research and innovation programme under the Marie Skłodowska-Curie grant agreement No 101067043.  
Claude was used to assist with language revision and the verification of the analysis workflow; all scientific decisions, validation of the results, and responsibility for the final manuscript remain with the authors.
\end{acknowledgements}

% Please note that we have included the references below in
% order to compile the document, but we ask you to:

\bibliographystyle{aa}
\bibliography{gal_inf}

% %%%%%%%%%%%%%%%%%%%%%%%%%%%%%%%%%%%%%%%%%%%%%%%%%%%%%%%%%%%%%%
% Example below of non-structurated natbib references  
% To use the v8.3 macros with this form of composition of bibliography,
% the option "bibyear" should be added to the command line
% "\documentclass[bibyear]{aa}".
% %%%%%%%%%%%%%%%%%%%%%%%%%%%%%%%%%%%%%%%%%%%%%%%%%%%%%%%%%%%%%%

% \begin{thebibliography}{}

%   \bibitem[1966]{baker} Baker, N. 1966,
%       in Stellar Evolution,
%       ed.\ R. F. Stein,\& A. G. W. Cameron
%       (Plenum, New York) 333

%    \bibitem[1988]{balluch} Balluch, M. 1988,
%       A\&A, 200, 58

%    \bibitem[1980]{cox} Cox, J. P. 1980,
%       Theory of Stellar Pulsation
%       (Princeton University Press, Princeton) 165

%    \bibitem[1969]{cox69} Cox, A. N.,\& Stewart, J. N. 1969,
%       Academia Nauk, Scientific Information 15, 1

%    \bibitem[1980]{mizuno} Mizuno H. 1980,
%       Prog. Theor. Phys., 64, 544
   
%    \bibitem[1987]{tscharnuter} Tscharnuter W. M. 1987,
%       A\&A, 188, 55
  
%    \bibitem[1992]{terlevich} Terlevich, R. 1992, in ASP Conf. Ser. 31,
%       Relationships between Active Galactic Nuclei and Starburst Galaxies,
%       ed. A. V. Filippenko, 13

%    \bibitem[1980a]{yorke80a} Yorke, H. W. 1980a,
%       A\&A, 86, 286

%    \bibitem[1997]{zheng} Zheng, W., Davidsen, A. F., Tytler, D. \& Kriss, G. A.
%       1997, preprint
% \end{thebibliography}

%%%%%%%%%%%%%%%%%%%%%%%%%%%%%%%%%%%%%%%%%%%%%%%%%%%%%%%%%%%%%%%
% Appendices must be placed after   \end{thebibliography}
% They will be placed automatically on a new page.
%%%%%%%%%%%%%%%%%%%%%%%%%%%%%%%%%%%%%%%%%%%%%%%%%%%%%%%%%%%%%%%
\begin{appendix}
%%%%%%%%%%%%%%%%%%%%%%%%%%%%%%%%%%%%%%%%%%%%%%%%%%%%%%%%%%%%%%%
% In the PDF output, floats should be placed
% under their own appendix, not before the title, nor after the
% title of the next appendix.

% In short appendices, onecolumn floats (\figure*
% or \table*) will generate a blank page.
% To prevent this behaviour, a few examples are provided here. 

% In case you have a lot of floating objects for little text and the 
% LaTeX engine moves the floats away from their context, the command
% \FloatBarrier of the “placeins” package will empty the
% float buffer and place all stored floats in the continuity.

% If you still encounter problems with wide floats placement,
% just use the onecolumn environment throughout the appendices.
%%%%%%%%%%%%%%%%%%%%%%%%%%%%%%%%%%%%%%%%%%%%%%%%%%%%%%%%%%%%%%%

%____________________________________________________________
%       Wide floats at the start of an appendix: first method
%-------------------------------------------------------------
% To prevent a blank page after the start of an appendix:
% - Switch to one \onecolumn first
% - Declare the section title
% - Declare the onecolumn float with the parameter [ht!]
% - Revert to \twocolumn at the end of the section

\section{Effect of the halo mass function}
\label{sec:appendixA}
\begin{figure}[h!]
    \centering
    \includegraphics[width=1.0\linewidth]{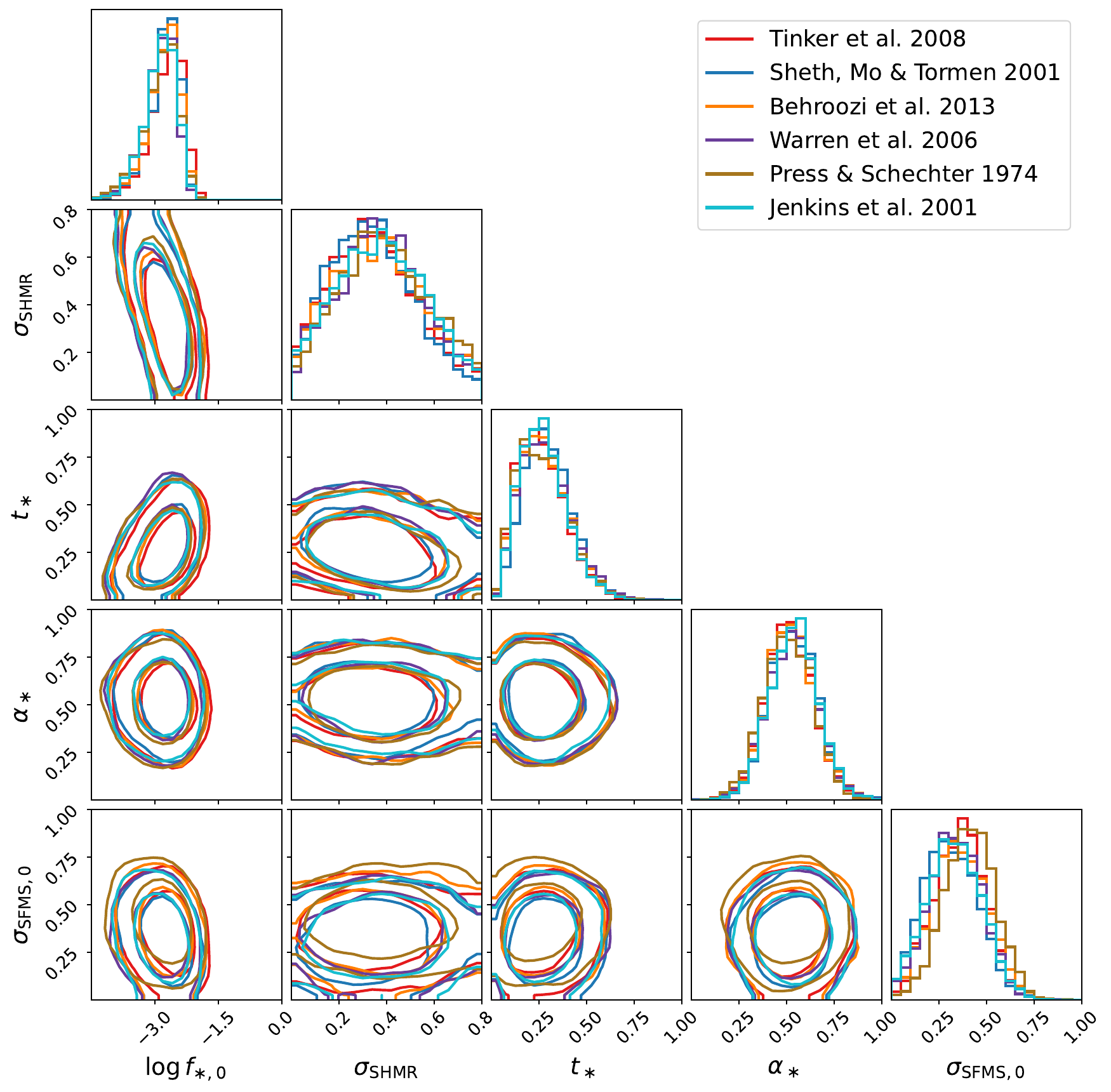}
    \caption{Posteriors (68\% and 95\% credible regions) of the parameters most affected by the choice of HMF, for the six HMFs listed in the text, from the UVLF + ACF inference. The differences between HMFs are smaller than the width of each posterior, which is set by the observational uncertainties.}
    \label{fig:change_hmf}
\end{figure}

In Fig.~\ref{fig:change_hmf}, we show how the choice of HMF affects the inferred parameters, using the same UVLF + ACF likelihood as in Sect.~\ref{sec:results}. We compare the HMFs of \citet{Tinker2008}, \citet{Sheth2001}, \citet{Behroozi2013}, \citet{Warren2006}, \citet{Press1974} and \citet{Jenkins2001}, and show the parameters most affected by this choice.

The choice of HMF shifts the posterior medians by at most $0.5\sigma$, i.e., by less than the posterior widths. Although the HMF is uncertain at high redshifts \citep{OBrennan2024}, this uncertainty is thus subdominant to that from the measurement errors. As future JWST observations map the UV LF more precisely, an accurate HMF will become more important.

\section{Inference using other UV luminosity function estimates}
\label{sec:app_inf_uvlf_others}
\begin{figure}
    \centering
    \includegraphics[width=1.0\linewidth]{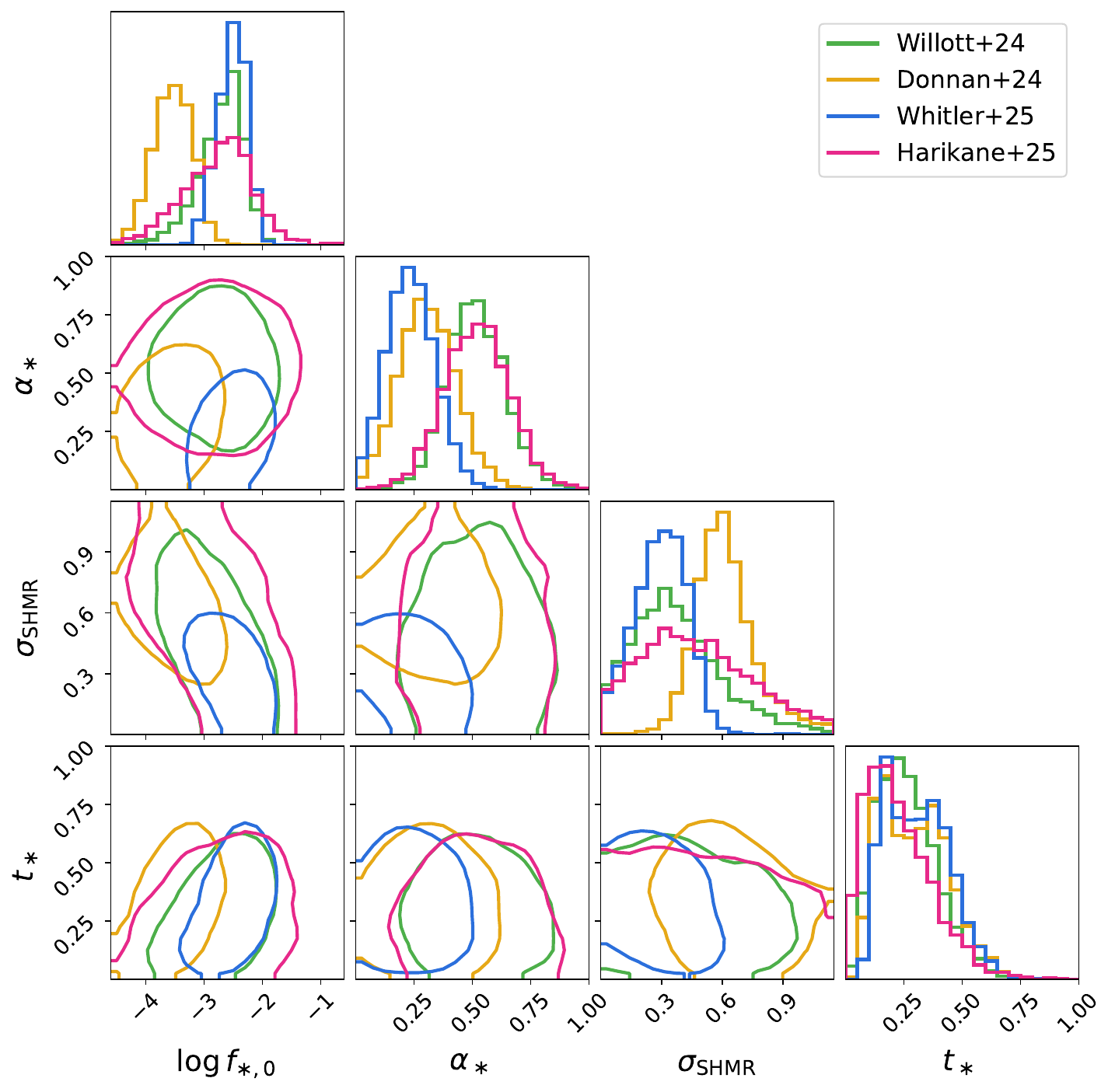}
    \caption{Posteriors of four model parameters inferred from the faint-end ($M_{\rm UV}>-20$) UV LFs of \citet{Willott2024}, \citet{Donnan2024}, \citet{Whitler2025} and \citet{Harikane2025}, each combined with the ACFs; contours show the 95\% credible regions.}
    \label{fig:uvlf_diff_inf}
\end{figure}
\begin{figure*}
    \centering
    \includegraphics[width=1.0\linewidth]{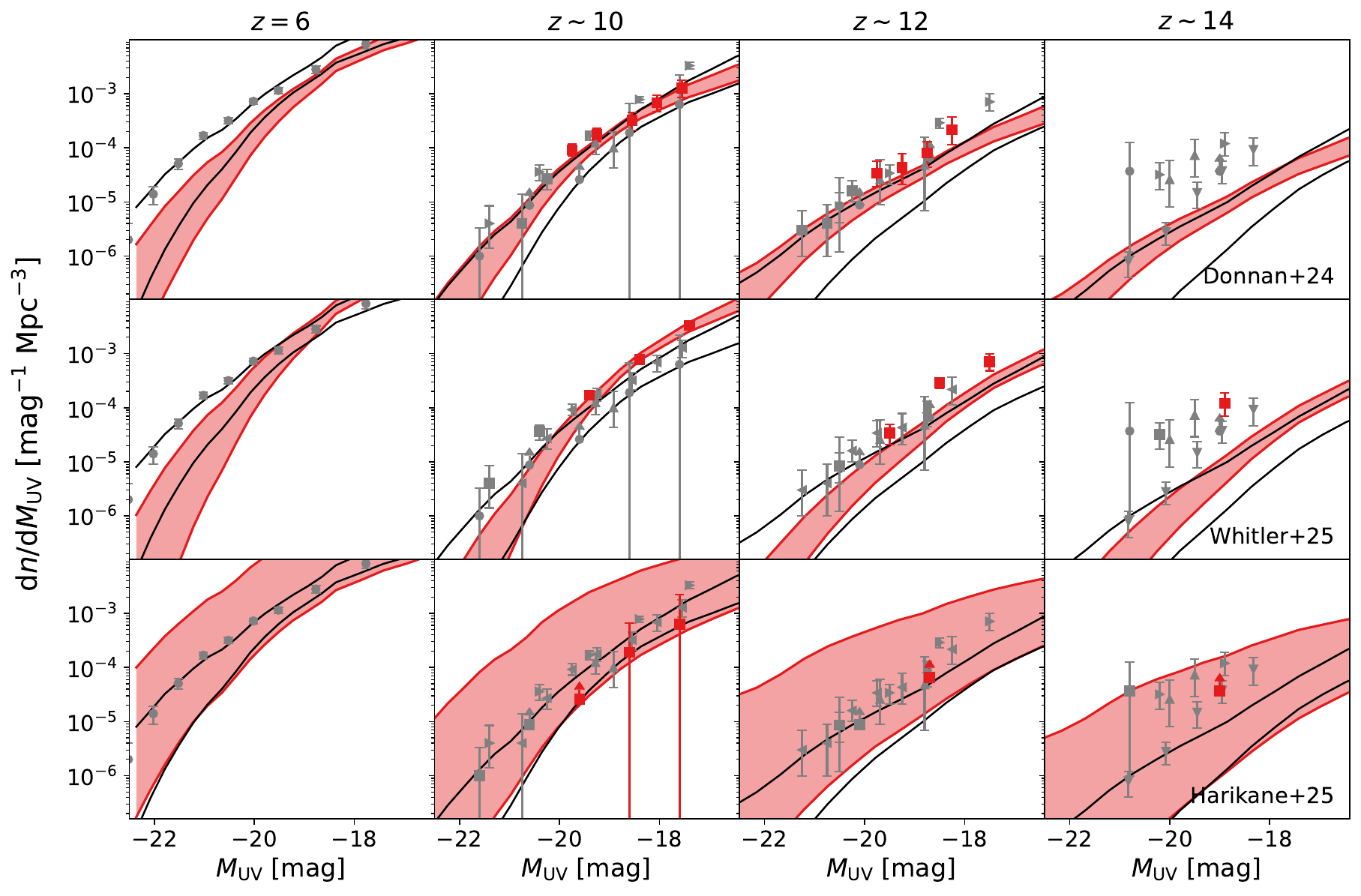}
    \caption{Posterior predictive UV LFs of the inferences in Fig.~\ref{fig:uvlf_diff_inf} that use the UV LFs of \citet{Donnan2024} (top), \citet{Whitler2025} (middle) and \citet{Harikane2025} (bottom), each combined with the ACFs. Red bands show the $95\%$ credible intervals, and black lines those of the fiducial UVLF + ACF inference (Fig.~\ref{fig:uvlf_post}). Columns show $z=6$, which is not fitted, and the bins of each dataset nearest to $z=10$, $12$ and $14$ ($z=14$ for \citet{Donnan2024} is an extrapolation as there is no estimate there); the black lines are at $z=6$, $10.23$, $12.5$ and $14$. Red squares show the data used in the likelihood of each row ($M_{\rm UV}>-20$; brighter points of the same dataset in gray). Other estimates are shown in gray, with the markers of Fig.~\ref{fig:uvlf_post} and arrows for lower limits. The figure assumes the same dust attenuation model as in Fig.~\ref{fig:uvlf_post}.}
    \label{fig:uvlf_diff_uvlf}
\end{figure*}

The UV LF estimates of different groups, based on different fields and pipelines, are in tension with each other, especially at the highest redshifts (Sect.~\ref{sec:uvlf}). In Fig.~\ref{fig:uvlf_diff_inf}, we show the posteriors of four parameters inferred, in turn, from the faint-end ($M_{\rm UV}>-20$) UV LFs of \citet{Willott2024}, \citet{Donnan2024}, \citet{Whitler2025} and \citet{Harikane2025}, each combined with the ACFs. The posteriors differ for three of the four parameters: \citet{Donnan2024} prefers a lower normalization, $\log f_{\ast,0}=-3.5\pm0.3$ ($68\%$ credible interval), about $1$\,dex below the other three estimates ($-2.5$ to $-2.7$), and $\alpha_\ast$ ranges from $\approx0.2$ \citep{Whitler2025} and $\approx0.3$ \citep{Donnan2024} to $\approx0.5$ \citep{Willott2024, Harikane2025}. \citet{Donnan2024} also prefers a large SHMR scatter, $\sigma_{\rm SHMR}=0.60^{+0.16}_{-0.12}$, whereas the other three favor smaller values (medians $0.30$--$0.46$). By contrast, the SFMS timescale $t_\ast$ is consistent across all four estimates, with posteriors peaking at $t_\ast\approx0.15$--$0.2$. Figure~\ref{fig:uvlf_diff_uvlf} shows the corresponding posterior predictive UV LFs. Without dust attenuation, which is not applied in the likelihood, each inference reproduces its own data at $z\approx9$--$10$ to within $0.05$\,dex. At higher redshifts, however, the faint-end points of \citet{Donnan2024} at $z=11$ (not shown) and $12.5$ and of \citet{Whitler2025} at $z=12.8$ lie $1.4$--$3.2$ times above the $95\%$ credible intervals of their own inferences, and the \citet{Whitler2025} point at $z=14.3$ about $10$ times above: these UV LFs decline with redshift more slowly than the model predicts. The same two inferences fall below the bright end of the HST UV LF at $z=6$, which is not fitted: the points at $M_{\rm UV}<-20$ lie up to $4$ times above the upper edge of the $95\%$ interval. The \citet{Harikane2025} data, mostly lower limits, constrain the model least; its predictions are the broadest and include both the HST data at $z=6$ and the \citet{McLeod2026} estimates at $z\sim14$. Before making strong claims, a consistent treatment between different JWST fields is needed \citep[see][]{McLeod2026}. 

\section{Response of the observables to each parameter}
\label{appendix:par_sweep}
\begin{figure*}
    \centering
    \includegraphics[width=0.9\linewidth]{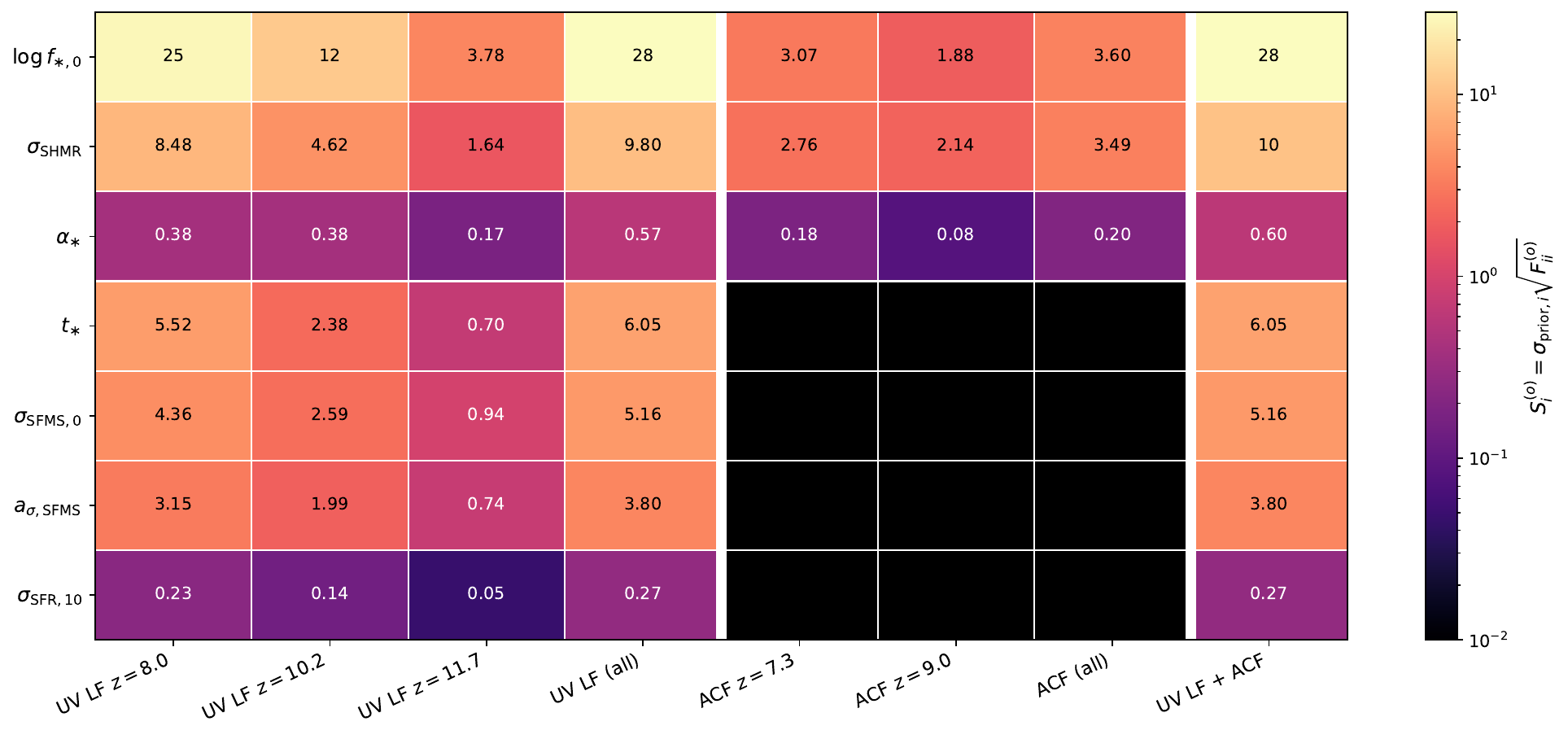}
    \caption{Same as Fig.~\ref{fig:sensitivity}, but for the data used in the inference: the \citet{Willott2024} UV LFs and the \citet{Paquereau2025} ACFs with their reported uncertainties, per redshift bin and combined. Black cells indicate that the observable does not depend on the parameter.}
    \label{fig:sensitivity_data}
\end{figure*}
\begin{figure*}
    \centering
    \includegraphics[width=1.0\linewidth]{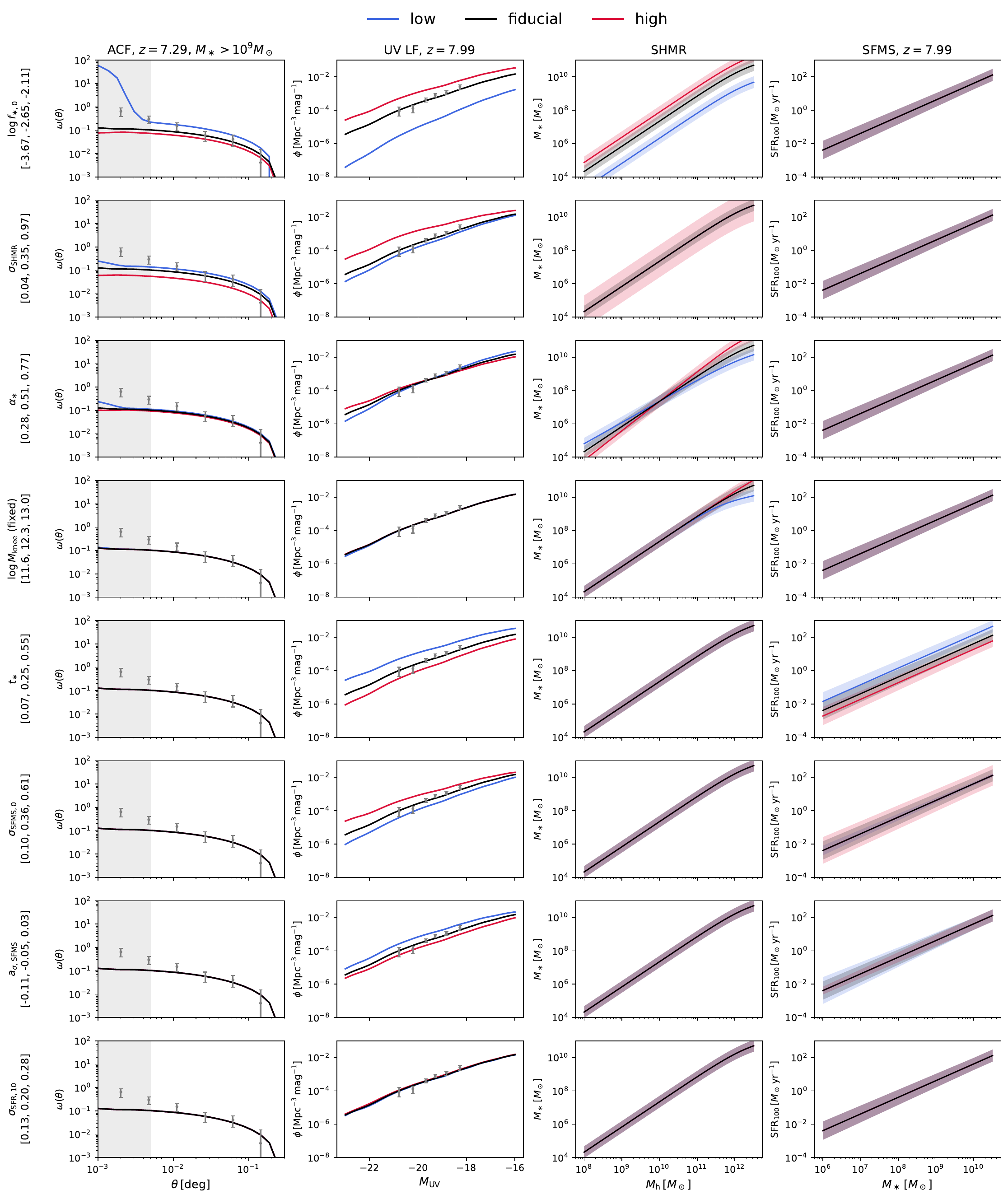}
    \caption{Response of the model predictions to one parameter at a time. In each row, one parameter takes the 2.5th percentile (blue), median (black) and 97.5th percentile (red) of the UVLF + ACF posterior, with the others held at their medians; the values are given on the left. $M_{\rm knee}$ is fixed in the inference, so its row varies it by $\pm0.7$\,dex. Columns show the ACF ($z=7.29$, $M_\ast>10^9\,M_\odot$), the UV LF ($z=7.99$), the SHMR and the SFMS ($z=7.99$). Bands show the $1\sigma$ scatter. Gray points show the \citet{Paquereau2025} ACF and the \citet{Willott2024} UV LF; the gray area marks ACF scales excluded from the fit ($\theta<0.005$\,deg).}
    \label{fig:full_parameter_sweep}
\end{figure*}
Figure~\ref{fig:sensitivity_data} shows the sensitivity $S^{(o)}_i$ of Sect.~\ref{sec:sensitivity} for the data used in the inference, i.e., with the reported uncertainties of the \citet{Willott2024} UV LFs and the \citet{Paquereau2025} ACFs. The UV LFs carry most of the information on $\log f_{\ast,0}$ ($S\approx28$), $\sigma_{\rm SHMR}$ ($\approx10$), $t_\ast$ ($\approx6$), $\sigma_{\rm SFMS,0}$ ($\approx5$) and $a_{\sigma,\rm SFMS}$ ($\approx4$), mostly from the $z\approx8$ bin, whereas the ACFs inform only $\log f_{\ast,0}$ and $\sigma_{\rm SHMR}$ ($S\approx3.5$ each). Neither constrains $\alpha_\ast$ ($S\approx0.6$) or $\sigma_{\rm SFR,10}$ ($S\approx0.3$) beyond the prior.

Figure~\ref{fig:full_parameter_sweep} shows how the UV LF, ACF, SHMR and SFMS respond to each parameter of Sect.~\ref{sec:galfor} across its $95\%$ posterior range.

Beyond the aggregate sensitivities summarized in Fig.~\ref{fig:sensitivity}, the parameter sweep reveals two features not captured by a single squared derivative per parameter–observable pair. First, several parameters move the ACF and UV LF in opposite directions: increasing $f_{\ast,0}$ or $\sigma_{\rm SHMR}$ raises the bright-end UV LF while lowering the ACF amplitude (Sect.~\ref{sec:sensitivity}). This mechanism, in reverse, produces the ACF overprediction in Fig.~\ref{fig:ang_elp}. Because the ACF weights the two parameters differently than the UV LF, it narrows the $f_{\ast,0}$--$\sigma_{\rm SHMR}$ degeneracy. Second, neither the ACF nor the UV LF responds visibly to $M_{\rm knee}$ over the sampled range, which lies above the halo masses probed by the data -- directly explaining why $M_{\rm knee}$ remains essentially unconstrained and is fixed in our fiducial inference (Sect.~\ref{sec:shmr}).

\section{Full posterior}
\label{appendixB}

\begin{figure*}
    \centering
    \includegraphics[width=0.9\linewidth]{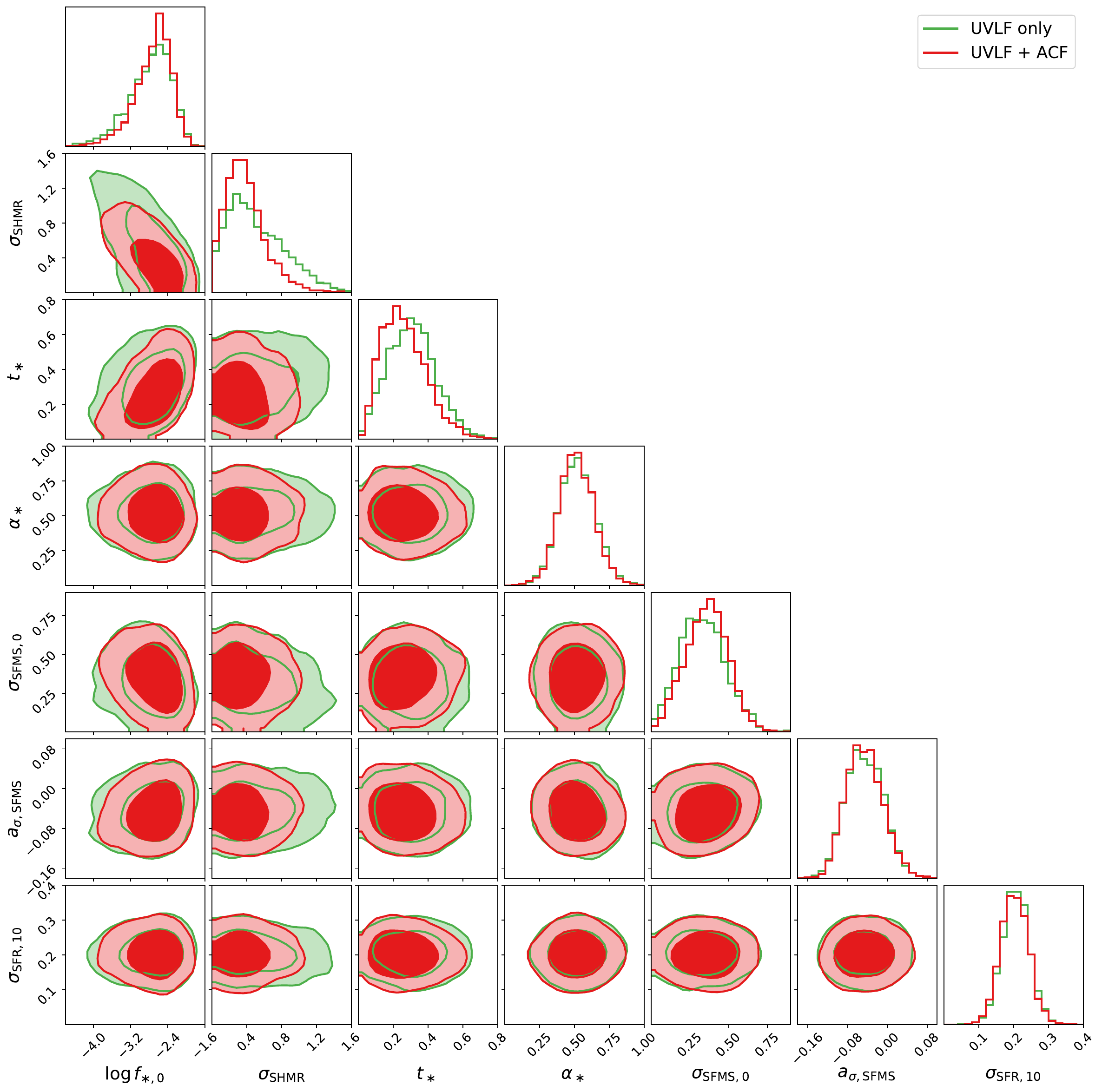}
    \caption{Same as Figs.~\ref{fig:corner_SHMR} and \ref{fig:sfms_corner}, but for the full parameter set; contours show the 68\% and 95\% credible regions.}
    \label{fig:full_corner}
\end{figure*}

Figure~\ref{fig:full_corner} shows the full posterior of both inferences. In addition to Figs.~\ref{fig:corner_SHMR} and~\ref{fig:sfms_corner}, it shows the covariances between the SHMR and SFMS parameters. Table~\ref{tab:params} lists the prior and posterior of each parameter.

\begin{table}
\caption{Prior and posterior of the model parameters. Priors are Gaussian (mean $\pm$ standard deviation), truncated to the given range; posteriors are the medians and $68\%$ credible intervals of the UVLF + ACF inference.}
\label{tab:params}
\centering
\small
\renewcommand{\arraystretch}{1.3}
\begin{tabular}{@{}llll@{}}
\hline\hline
Parameter & Prior & Range & Posterior \\
\hline
$\log f_{\ast,0}$ & $-2.51\pm1.55$ & $[-6, 1]$ & $-2.65^{+0.30}_{-0.45}$ \\
$\alpha_\ast$ & $0.51\pm0.16$ & $[0, 2]$ & $0.51^{+0.13}_{-0.12}$ \\
$\sigma_{\rm SHMR}$ & $0.31\pm0.60$ & $[0, 2]$ & $0.35^{+0.25}_{-0.19}$ \\
$t_\ast$ & $0.18\pm0.21$ & $[0, 1]$ & $0.25^{+0.14}_{-0.12}$ \\
$\sigma_{\rm SFMS,0}$ & $0.35\pm0.19$ & $[0, 1.5]$ & $0.36^{+0.13}_{-0.14}$ \\
$a_{\sigma,\rm SFMS}$ & $-0.045\pm0.048$ & $[-1, 0.5]$ & $-0.047^{+0.039}_{-0.035}$ \\
$\sigma_{\rm SFR,10}$ & $0.20\pm0.045$ & $[0, 0.5]$ & $0.20\pm0.04$ \\
$\log(M_{\rm knee}/M_\odot)$ & \multicolumn{3}{l}{$12.3$ (fixed)} \\
\hline
\end{tabular}
\end{table}

\FloatBarrier %\usepackage{placeins}
\clearpage

\end{appendix}
\end{document}